\documentclass[11pt]{article}
\pdfoutput=1
\usepackage[margin=1.0in]{geometry}
\usepackage{amsmath,amssymb,graphicx}
\usepackage[colorlinks=true, urlcolor=blue, linkcolor=blue]{hyperref}
\usepackage{braket}
\usepackage{csquotes}
\usepackage{float}

\usepackage{tikzducks}

\usepackage{framed}

\usepackage[font=small,labelfont=bf]{caption}
\usepackage{cite}

\newcommand{\be}{\begin{equation}}
\newcommand{\ee}{\end{equation}}

\usepackage{xcolor}
\usepackage{bm}

    \title{\bf \large Lecture Notes for the \href{https://sites.psu.edu/cteq/cteqss2026/}{2026 CTEQ Summer School}\\ \ \\ \Large The two-dimensional electron Wigner crystal:\\  What's old and what's new?}
\author{Brian Skinner \\
{\small \color{gray} \texttt{skinner.352@osu.edu}}\\
{\small Ohio State University}}

\begin{document}

\maketitle

\begin{abstract}
A two dimensional electron system can exhibit a liquid-solid transition even at zero temperature due to the competition between quantum kinetic energy and repulsive Coulomb interactions. The solid phase, called the Wigner crystal (WC), is in some sense the oldest strongly-correlated phase of electrons, having been proposed in 1934. But it is only in the last few years that experiments have convincingly demonstrated the existence of electronic WCs in solid materials without magnetic field. These experiments have prompted a flurry of new questions and new understanding about the WC. Some of these questions are ``old-fashioned" and concern the fundamental nature of the WC and its quantum melting/freezing transition. Other questions are ``newfangled" and consider how the Wigner crystal is altered by the presence of Berry curvature. In these lecture notes I will review a smattering of old and new ideas about the Wigner crystal in the context of recent experiments. \\ 
This friendly duck will keep you company as you read: \resizebox{8mm}{8mm}{ \tikz\duck[crazyhair=gray, eyebrow=gray, glasses, book=\scalebox{0.5}{WC}];}
\end{abstract}

\thispagestyle{empty}

\tableofcontents

\newpage

\setcounter{page}{1}

\section{Old-fashioned physics: Wigner crystals in the jellium model}

\subsection{What is a Wigner crystal?}
\label{sec:whatis}

Let's start with a thought experiment. Imagine that you put a large number $n L^2$ of electrons in a large, two-dimensional\footnote{What does it mean for a box to be two-dimensional? In this case it doesn't mean that the third dimension $z$ doesn't exist, only that the thickness $L_z$ of the box along the $z$ direction is so small that the energy associated with confinement in the $z$ direction, $\sim \hbar^2/mz^2$, is much larger than any other energy scale, so that all electrons are locked in the lowest energy level of $z$ confinement.} %
square box of size $L$ and cool the system down to zero temperature ($n$ is the number of electrons per unit area). What happens?

This is actually kind of a trick question. The electrons are charged, so if you literally put the electrons in an empty box, you will have made an object with a huge net charge $Q = -e n L^2$. The resulting electric fields will be enormous, they will push all the electrons to the periphery of the box, and they will be associated with a massive Coulomb self-energy $\sim Q^2/L$.\footnote{I'm basically never going to write $4 \pi \varepsilon_0$ in these notes. I hope you get used to that quickly.} In other words, you will have made a ``Coulomb bomb'' -- if your box is $L = 10$\,cm in size and you add enough electrons for there to be 1 per square nm on average, then the Coulomb self-energy of the box is something like 200 kJ. That's about as much energy as a kg-sized C4 explosive.

So if you want to pose a sensible question, you have to stipulate that you \emph{also} add a neutralizing background of uniform positive charge to the box (charge concentration $+ e n$). This is often called the ``jellium model": one imagines that the electrons are sitting on an otherwise featureless plane of uniform background charge.

If you are reading these notes then you are probably an educated person who has taken some quantum mechanics, so you will think ``aha, now that the pesky Coulomb interaction is taken care of, this is basically just a Fermi gas problem: as I add electrons, they fill the kinetic energy levels one at a time before finally stopping at the Fermi energy, $E_F \sim \hbar^2 n / m$''.\footnote{I am also going to be doing a lot of dropping numeric factors and writing equations with ``$\sim$'' instead of ``$=$''. I hope you get used to that quickly also.} This is certainly an option on the table.

\begin{figure}[tb!]
	\centering
	\includegraphics [width = 0.8\textwidth]{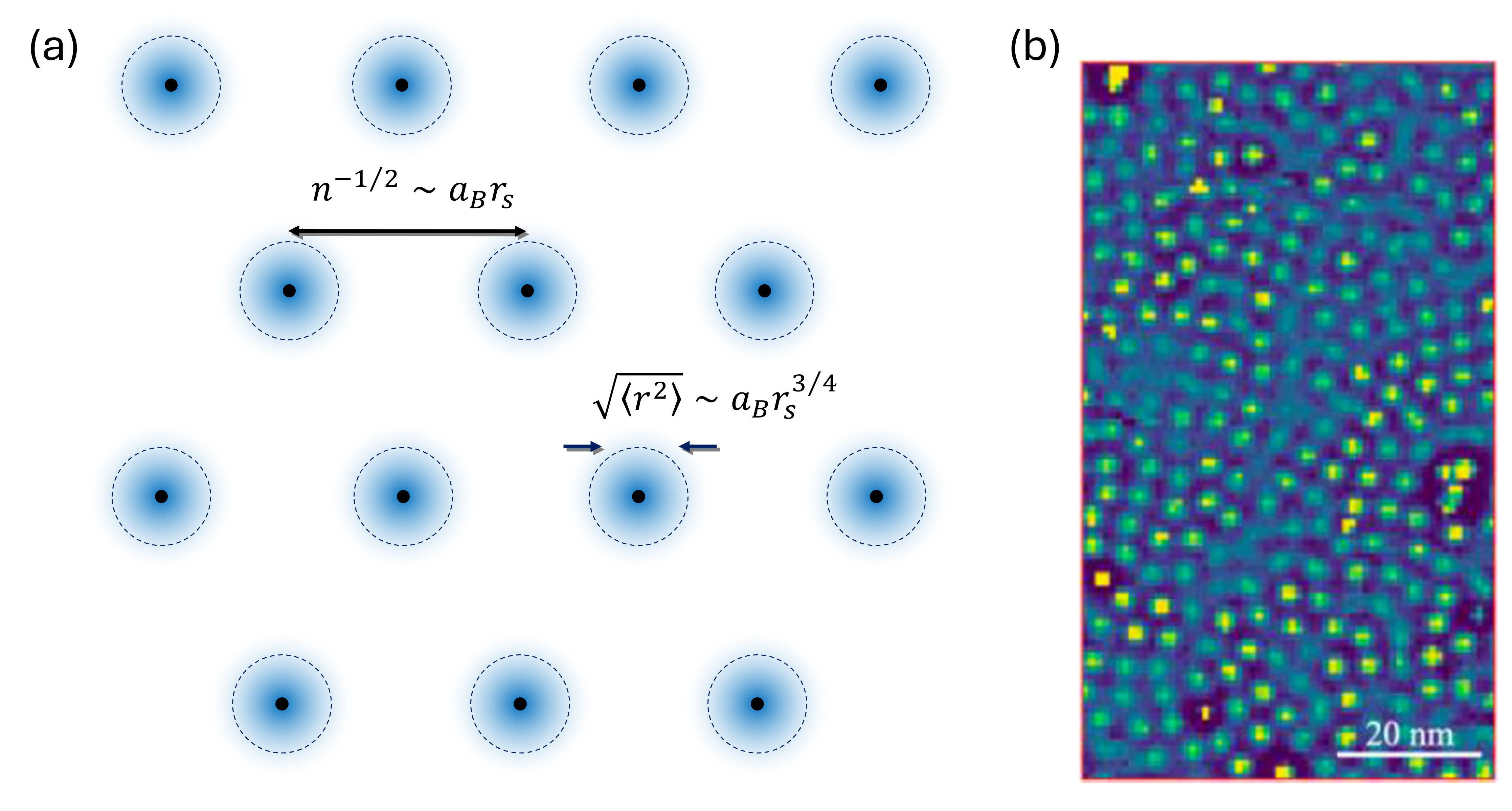}
	\caption{(a) A cartoon picture of a Wigner crystal, in which electron wave packets are arranged in a triangular lattice. (b) An experimental image of a Wigner crystal in bilayer MoSe$_2$, taken using scanning tunneling microscopy \cite{xiang2025imaging}.} 
	\label{fig:WCschematic}
\end{figure}

But now try to think back to a more primitive time in your life, before you learned any quantum mechanics, and consider how you would have responded to the question ``what do the electrons do at zero temperature?''. At this developmental stage, the only relevant physics you would have known is Coulomb's law, $E = e^2/r$, so you would have concluded that the electrons arrange themselves in such a way that they maximize their separation $r$ from each other (while still covering neutralizing the background charge in a more or less uniform way). This arrangement turns out to be a triangular lattice, as illustrated in Fig.~\ref{fig:WCschematic}. The only energy scale in the problem would be the Coulomb energy, $E_C \sim e^2 n^{1/2}$, since the spacing $r$ between neighboring electrons is $r \sim n^{-1/2}$.

The crucial idea of the Wigner crystal (WC), first proposed by Wigner in 1934 \cite{Wigner1934On}, is that there are situations where your more primitive, non-quantum-mechanics-knowing self is basically correct. These are situations where the Coulomb energy scale $E_C$ is much larger than the Fermi energy scale $E_F$; in these situations the electrons prefer to maximize their separation from each other (minimizing $E_C$) even if it leads to a higher quantum-mechanical kinetic energy.  The ratio between the two energy scales is called the ``interaction parameter''
\begin{equation} 
r_s = \frac{1}{\sqrt{\pi n a_B^2}} \sim \frac{E_C}{E_F},
\label{eq:rs}
\end{equation} 
where
\be 
 a_B = \frac{\hbar^2 \varepsilon}{m e^2}
 \ee 
 is the ``effective Bohr radius'', with $\varepsilon$ the dielectric constant and $m$ the electron mass. (That factor of $\pi$ in Eq.~(\ref{eq:rs}) is basically there for historical reasons: it comes from imagining that each electron occupies a circular territory of radius $r_s$ and area $\pi r_s^2 = 1/n$.) The WC phase corresponds to the limit $r_s \gg 1$ (as we shall see below, it ends up being more like $r_s > 30$). The usual Fermi gas corresponds to the limit $r_s \ll 1$ (although in practice even our most canonical ``weakly-interacting metals'' like copper or aluminum usually have $r_s$ in the range $1 - 5$).

It is worth pausing to make a brief commentary about units. In the jellium problem, the only relevant constants are $e^2/\varepsilon$, $\hbar$, and $m$. Setting all of these equal to unity gives ``Hartree units", which means that all distances are measured in units of the effective Bohr radius $a_B$ and all energies are in units of the Hartree energy $e^2/(\varepsilon a_B) = m e^4/(\varepsilon^2 \hbar^2)$. In atomic physics $a_B$ is very small (half an Angstrom) and the Hartree energy is very large ($26.4$\,eV). But in the solid state context the effective mass $m$ can be small and the dielectric constant can be large, so that $a_B$ is often as large as a few nm and the Hartree energy as small as a hundred meV or so. Thus, the interaction parameter $r_s$ can be understood as just the distance between electrons, measured in units of the natural length scale $a_B$.

A strange consequence of Eq.~(\ref{eq:rs}) is that the strongly interacting limit ($r_s \gg 1$) is the limit where electrons are far away from each other (low density). Electron-electron interactions are certainly weaker at larger $r_s$, since the Coulomb energy falls as $1/r$, but the Fermi energy drops off even faster, as $1/r^2$. One can therefore say that in the WC regime, \emph{all} energy scales are small. For this basic reason realizing a WC experimentally has been very difficult historically, and for a long time the only realization was electrons floating on the surface of liquid helium \cite{grimes1979evidence}. (And even in this context the WC was essentially classical, with the thermal energy $k_B T$ being higher than any quantum mechanical energy scale.) Another realization of WC physics came in the 1980s and 90s by putting the electrons in a strong magnetic field, which essentially killed the quantum kinetic energy by locking all electrons into Landau levels (see, e.g., \cite{lam1984liquid} for an early theory paper and \cite{tsui_direct_2024} for a cool recent experiment). In these notes I won't be talking about Landau levels or magnetic field and I will focus on the limit of zero temperature.

You might be wondering why the jellium model is at all relevant for electrons in real materials, since in real materials electrons live on a lattice of atoms rather than on a featureless uniform background. The answer is that if the effective Bohr radius $a_B$ associated with the material is much longer than the lattice constant $a_0$, then the electron wave packets associated with the WC (which we'll discuss in Sec.~\ref{sec:semiclassics}) are always effectively averaging over many atoms, so that, from the perspective of the electron, the background charge contained in the atomic nuclei looks pretty uniform. (Of course, the electron wave packets are actually made from the Bloch functions of the electron band, and both the Bloch functions and the band structure (including the effective mass) certainly know about the underlying atomic lattice.) In situations where $a_B \gg a_0$, the triangular lattice into which the electrons crystallize (called the ``Wigner lattice") has no relation with the underlying atomic lattice. In the opposite scenario where $a_B$ is of the same order of the lattice constant, there is no real notion of a WC, and the details of the microscopic atomic potential are always important for determining the electron ground state.

In 2D materials, there is also a way of continuously changing both the electron concentration and the ``jellium" density via electrostatic gating. In this setup one makes the 2D electron system (2DES) one half of a plane capacitor, while a flat piece of metal is the other half. Changing the gate voltage between the two halves of the capacitor injects electrons into the 2DES while depositing an equal and opposite amount of charge onto the surface of the gate electrode. Energetically, the energy of this capacitor is equal to the energy of the jellium model plus the energy of a uniform plane capacitor:
\begin{figure}[h!]
	\centering
	\includegraphics [width = 0.9\textwidth]{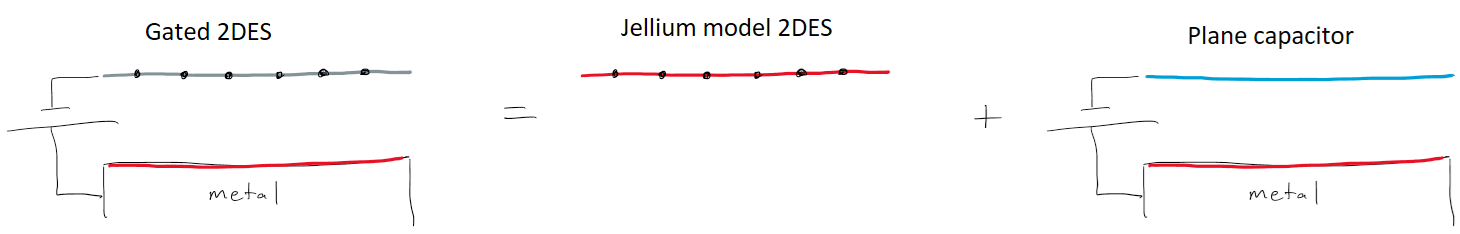}
	\caption{A gated 2DES is energetically equivalent to the jellium model 2DES plus the energy of a uniform plane capacitor.}
	\label{fig:capacitor}
\end{figure}

\noindent This pictorial equality is correct as long as the gate is not too close to the plane of the 2DES, as we discuss in Sec.~\ref{sec:phase_transitions}.

\subsection{Semiclassical considerations: collective modes and quantum corrections to the energy}
\label{sec:semiclassics}

So the lowest-order description of a WC is that it is basically a classical object,\footnote{even though it has been advertised as a ``quantum crystal'' in various recent news releases} in the sense that it arises in situations where the quantum mechanical energy is much smaller than the classical electrostatic energy. But of course everything about the WC must be consistent with the laws of quantum mechanics. So how do we describe the WC state in a way that is respectful of quantum mechanics? Said in a more formal way, how is the energy of the WC modified by the existence of quantum mechanical fluctuations?

There are two ways to answer this question: the honest way and the cheating way. The honest way involves carefully considering all the collective modes of the WC. Like any other crystal, the WC has phonon modes, both longitudinal (compression waves) and transverse (shear waves). The longitudinal waves are like the usual sound waves that exist in any solid or fluid, except for the fact that the electrons carry charge, which means that there is a long-ranged interaction between different regions of the sound wave, since a compression wave involves making regions of net charge (the compressed regions have a net negative charge and the rarefied regions have net positive charge). The long-ranged Coulomb interaction makes the longitudinal mode very ``stiff": at long wavelength the longitudinal sound wave has frequency $\omega \sim \sqrt{e^2 n / (m \varepsilon)} \times q^{1/2}$, where $q$ is the wave vector, just like in the plasmon mode of a 2D charged fluid. The transverse (shear) mode, on the other hand, is unique to the solid, and it arises from the finite shear modulus (e.g., the fact that there is an energy barrier associated with sliding two planes of the Wigner lattice past each other). This shear mode is much softer, having $\omega = v_\perp q$, with a small sound velocity $v_\perp \approx 0.5 \sqrt{e^2 n^{1/2}/m\varepsilon}$ \cite{fisher1979defects} owing to the small shear modulus of the WC. One can view the shear mode as the Goldstone mode associated with translational symmetry breaking.

The honest way to calculate the quantum mechanical correction to the energy of the WC is to calculate the full dispersion relation $\omega(q)$ for each of these two phonon modes (e.g., for every wave vector $\vec{q}$ in the Brillouin zone of the Wigner lattice) and then assign one quantum of (ground state) energy $\hbar \omega(\vec{q})/2$ to every possible mode $\vec{q}$. If you do this somewhat laborious calculation (first explained in Ref.~\cite{bonsall1977some}), you arrive at an energy
\begin{equation}
	E \stackrel{\text{honest}}{=} \frac{e^2}{\varepsilon a_B} \left( - \frac{1.11}{r_s} + \frac{0.81}{r_s^{3/2}} + \mathcal{O}\left( r_s^{-2} \right) \right).
	\label{eq:honest}
\end{equation}  
(Remember that the WC phase exists at large $r_s \gg 1$, so increasing powers of $1/r_s$ are increasingly less important.)

The leading order term of Eq.~(\ref{eq:rs}) is completely classical in nature: it corresponds to the energy per electron of a triangular lattice of point charges in a neutralizing background. Written in terms of density, this leading-order energy is
\begin{equation}
	E \simeq - 1.9 \frac{e^2 n^{1/2}}{\varepsilon}.
	\label{eq:Eclassical}
\end{equation}
Notice that the energy is \emph{negative}, and this is because we have a neutralizing background charge. Since electrons are doing a good job of avoiding each other, they experience more attraction to the positive background than repulsion from each other, and consequently their total electrostatic energy is negative. Indeed, one can make a good approximation of the energy of a WC by calculating the total electrostatic energy of a single, negative point charge sitting at the center of a single unit cell of positive background charge -- this was Wigner's original calculation \cite{Wigner1934On}. As we shall see in Sec.~\ref{sec:WCsignatures}, the negative sign in Eq.~(\ref{eq:Eclassical}) has an important experimental implication.

Life is short, and summer school courses are even shorter, so let's talk now about the cheating way to incorporate quantum mechanics into our description of the WC. The cheating way is to think about each electron in isolation as something like a little harmonic oscillator sitting in the confining potential created by its neighbors. That is, we treat all electrons as point charges except for one -- say, the electron at the origin, assuming that the origin is one of the sites of the Wigner lattice --; we calculate the Coulomb potential $V(r)$ created by all these other electrons; and we expand $V(r)$ to second order in the distance $r$ from the origin. This is basically an ``Einstein phonon" description, in analogy to Einstein's classic calculation of the specific heat of a solid by imagining each atom as an independent harmonic oscillator \cite{einstein1907plancksche}.

As is so often the case in life, the cheating way is much easier than the honest way. In this case, it still involves an infinite sum (to calculate the confining potential), but you can tell right away just by dimensional analysis that you are going to get $V(r) \sim e^2 n^{3/2} r^{2}/\varepsilon$.\footnote{Can I stop writing $\varepsilon$ now? From now on, ``$e^2$'' means $e^2/\varepsilon$, or $e^2/(4 \pi \varepsilon_0 \varepsilon)$ if you are the kind of pedant who insists on thinking in SI units.} Now the first quantum correction to the energy is just the ground state of the 2D harmonic oscillator: $\hbar \omega$, where $V(r) \equiv \frac12 m \omega^2 r^2$. A little bit of algebra gives $E_Q \sim \sqrt{\hbar^2 e^2 n^{3/2}/m}$, which is equivalent (up to the numeric factor) to the second term in Eq.~(\ref{eq:honest}). And this term is positive, of course: allowing for quantum mechanical fluctuations of the electron positions around the sites of the Wigner lattice increases both their kinetic energy (from zero) and their interaction energy.

In fact, the cheating way is much more accurate quantitatively than it has any right to be. It gives
\begin{equation}
E \stackrel{\text{cheating}}{=} \frac{e^2}{\epsilon a_B} \left(- \frac{1.11}{r_s} + \frac{0.89}{r_s^{3/2}} + \mathcal{O}\left( r_s^{-2} \right) \right),
\end{equation} 
which [comparing to Eq.~(\ref{eq:honest})] means that it gets the quantum correction right to within about $8\%$. The associated size of the electron wave packet is 
\be 
\sqrt{\langle r^2 \rangle} \sim a_B r_s^{3/4},
\label{eq:wavepacketw}
\ee 
so, as mentioned in the previous section, the wave packet is larger than $a_B$ in the WC phase ($r_s \gg 1$) and much smaller than the distance $n^{-1/2} \sim a_B r_s$ between electrons. 

Even more impressively, you can use the cheating way to make an estimate of the location of the WC - Fermi liquid (FL) phase boundary. As we increase the electron density from zero (and thereby reduce the value of $r_s$ from infinity), the relative importance of quantum fluctuations becomes more important, and at some critical value of $r_s = r_{s,c}$ these fluctuations are big enough to induce a solid-to-liquid phase transition.
Accurately calculating the value of $r_{s,c}$ is a very difficult problem. Since analytical approaches generally require some kind of perturbative approach where either interactions (which are relatively small in the liquid phase) or kinetic energy (which is relatively small in the solid phase) are treated as a perturbation, there are not really any analytical approaches available to us when the two energy scales are nearly exactly balanced. Instead one must resort to precision numerics like quantum Monte Carlo, which give a critical $r_s$ somewhere in the range $r_s = 30 - 40$ \cite{drummond2009phase, azadi2024quantum}.

\begin{figure}[H]
	\centering
	\includegraphics [width = 0.5\textwidth]{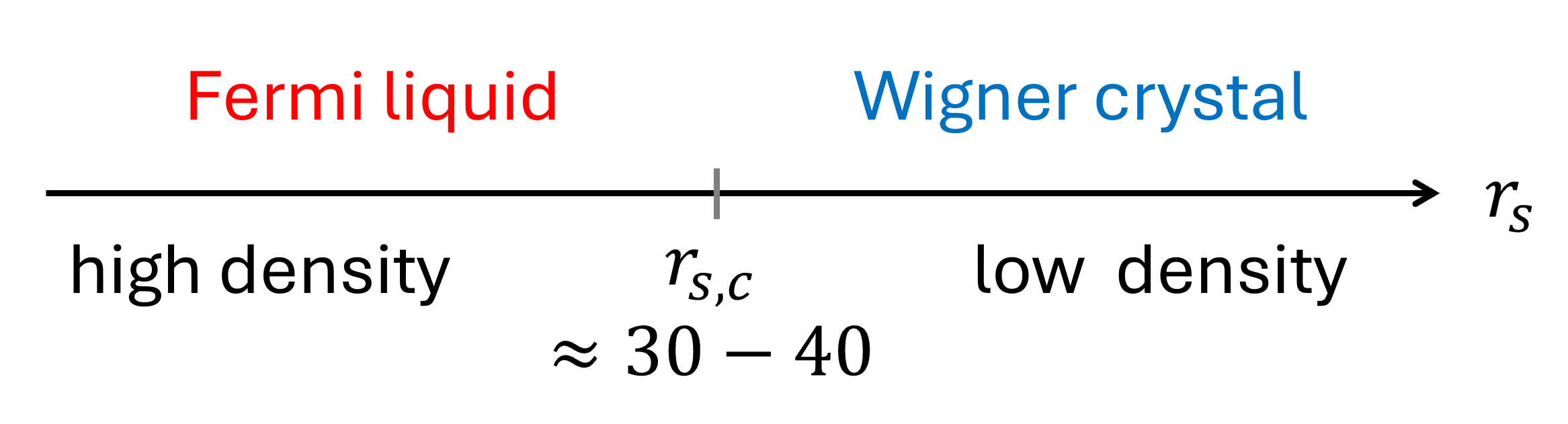}
	\caption{Schematic phase diagram of the 2DES.} 
	\label{fig:SimplePD}
\end{figure}

But the ``Einstein phonon" picture offers a remarkably simple way of thinking about the quantum melting transition by using the semi-empirical ``Lindemann criterion''. The Lindemann criterion says that a crystal (of any kind) melts when the root-mean square fluctuations of a given atom about its equilibrium position become larger than a critical fraction $\eta_c$ of the lattice constant. This criterion applies equally well to thermal (classical) and quantum fluctuations, and in 2D its value is almost universally in the range $\eta_c = 0.21 - 0.25$ \cite{babadi2013universal}. Within the Einstein phonon picture, the root mean square fluctuation of the electron is given exactly by the width of the (Gaussian) ground state wave function of the harmonic oscillator, i.e., by Eq.~(\ref{eq:wavepacketw}). Equating $\sqrt{\langle r^2 \rangle}$ to $\eta_c$ times the lattice constant of the Wigner lattice, and using $\eta_c = 0.23$, gives a critical $r_s = 34$. By contrast, Hartree-Fock calculations (incorrectly) predict a critical $r_s$ in the range $1 - 4$ \cite{trail2003unrestricted, bernu2011hartree}, which means that they overpredict the stability of the WC phase by over $100$ times in the electron density.

Of course, the cheating way has no business being \emph{quite} this good, but its remarkable track record has emboldened me personally to use it pretty liberally, at least for semi-quantitative understanding.

\subsection{What is the nature of the liquid-solid phase transition?}
\label{sec:phase_transitions}

Okay, the question ``how does a WC melt into a liquid?'' is a big can of worms. Let's not open it all the way, but perhaps we can lift the lid just enough to admire how unexpectedly interesting the worms are.

At first sight, the electronic liquid-solid transition is a first order transition, just like ice melting into water. (It also happens that electrons are ``weird" in the same way that water is, in the sense that the solid phase is less dense than the liquid.) If you calculate the free energy $f$ per unit area of both phases as a function of the electron density, there is a point where the two curves cross, which is the nominal location of the liquid-solid phase transition. This crossing implies a discontinuity of the derivative $df/dn$, or in other words a discontinuity in the chemical potential $\mu$.

\begin{figure}[htb!]
	\centering
	\includegraphics [width = 0.8\textwidth]{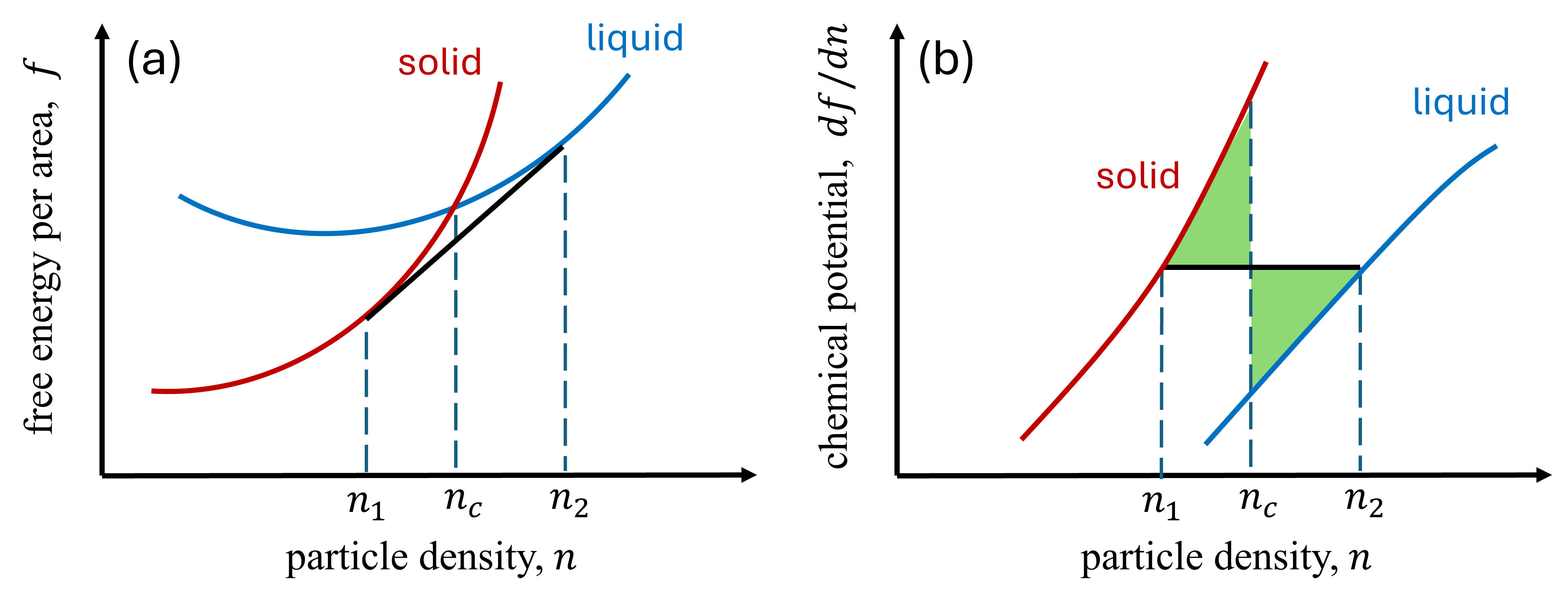}
	\caption{(a) In the Maxwell construction for phase coexistence, a system can lower its free energy relative to either the uniform solid or uniform liquid phases by making a phase mixture in which solid domains having density $n_1$ coexist with liquid domains having density $n_2$. The total free energy follows the black tangent line. (b) In terms of the chemical potential $df/dn$, the Maxwell construction is defined by the solid and liquid phases coexisting at the same chemical potential, while the two indicated green areas are equal.} 
	\label{fig:Maxwell}
\end{figure}

Any time a system exhibits a discontinuity in the chemical potential, there is always a way to lower the energy of the system near the transition (for a fixed particle number) by forming a mixture of the two phases, each at a different value of the density.  Maxwell pointed this out in 1875, and codified the rules for how phase mixtures work in what is now known as the ``Maxwell construction". The rules of the Maxwell construction are depicted in Fig.~\ref{fig:Maxwell}.

So a natural thing to expect is that in the vicinity of the WC-FL transition, there will be a window of average electron density in which large chunks of electron solid float in the electron liquid, like the frozen surface of a Pennsylvania lake in January.

But it turns out that macroscopic phase separation is not allowed for the WC-FL transition. If you propose this scenario, then you have slipped up and made a Coulomb bomb again. Since the liquid and the solid regions exist at different densities, and since the jellium background is uniform, a macroscopic region of size $\sim L$ has a huge Coulomb self energy $\sim (e\, \delta n \, L^2)^2/L$ that diverges with $L$, where $\delta n$ is the difference in density between the liquid and solid phases.

\begin{figure}[tb!]
	\centering
	\includegraphics [width = 0.8\textwidth]{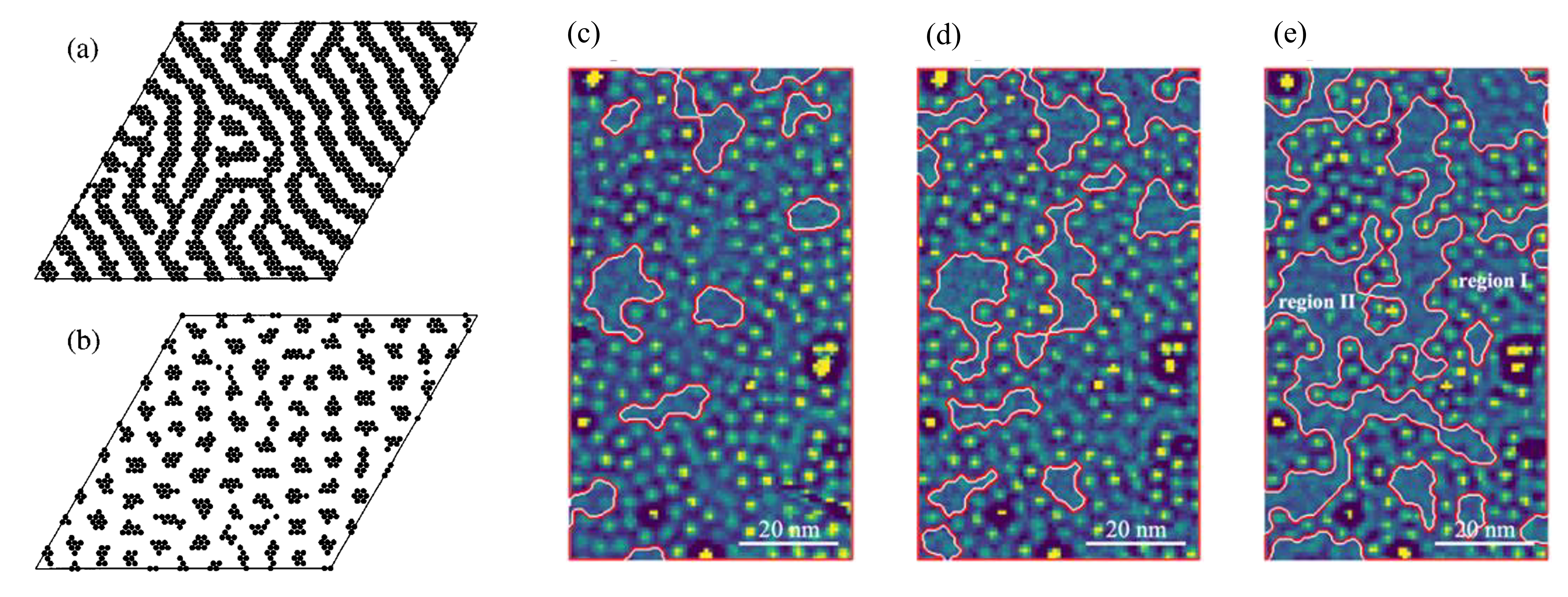}
	\caption{(a)-(b) Schematic of a microemulsion-type state (from Ref.~\cite{fogler1996ground}), where solid and liquid phases are intermixed. The two phases form stripes (a) when they are present in similar amounts, and bubbles (b) of the minority phase otherwise.  (c)-(e) show scanning tunneling microscopy images (from Ref.~\cite{xiang2025imaging}) of the same region of the sample during the WC-FL transition, with electron density increasing to the right. Regions outlined in red are ``liquid lakes'' emerging within the WC.} 
	\label{fig:microemulsions}
\end{figure}

Instead, it was pointed out by Spivak and Kivelson \cite{Kivelson_quantization_1986, spivak2003phase, spivak2004phases} that the electronic liquid-solid transition must proceed through a sequence of ``microemulsions", in which the liquid and solid phases are densely mixed in mesoscopic domains. As the overall density is varied through the ``microemulsion'' window, the sizes and shapes of these microemulsions vary in interesting ways, as shown in Fig.~\ref{fig:microemulsions}. [You might object to my use of the word ``must'' in the first sentence of this paragraph, and think that perhaps the system could just pass directly from a uniform WC to a uniform FL as a function of density. But there is a nice proof that this scenario is not allowed either. I sketch this proof in Sec.~\ref{sec:Kivelsonproof}.]

In the disorder-free limit, these microemulsion domains are spontaneously formed in a way that minimizes the total energy (including the surface tension between the phases), and they generally look like alternating stripes of two similarly-abundant phases or droplets of the minority phase embedded in the majority phase. Experiments have observed the formation of something like microemulsions in the vicinity of the WC-FL transition \cite{xiang2025imaging, sung2025electronic}, but it's likely that these microemulsions are driven by disorder rather than by interactions \cite{joy2023upper, joy2026disorder}. When disorder is present in the form of stray charges (either in the plane with the electrons or embedded in the substrate), it produces a continuous random electric potential that spatially modulates the electron density, so that certain spatial regions pass through the phase transition earlier than others. But even if there is only short-ranged disorder (like isovalent substitutions of one atom for another), you can still get phase mixing because disorder stabilizes the WC phase over the FL phase. The basic idea is that, by virtue of having pre-made electron wave packets, the WC phase is able to accommodate disorder (by slight positional shifts of these wave packets) more easily than the FL phase \cite{joy2026disorder, hammam2025disorder}. So regions that by chance have higher defect density will see a persistence of the WC phase to higher density, leading to random, disorder-induced spatial mixing of the two phases.

There is one other complication that arises in the gate-controlled 2DES. When a metal is present at some distance $d$ from the plane of the 2DES, an electron generally induces an image charge in the surface of the metal. In fact, the ``jellium background'' that is supposed to be present in the metal is not really uniform at all -- it is exactly a set of positive image charges \cite{skinner2010anomalously}.  When the gate is far away, the electric potential created by these image charges is basically identical to that of a uniformly charged plane. But when the gate is closer than the distance $n^{-1/2}$ between electrons, it provides strong \emph{screening} of the electron-electron interaction. Specifically, the electron-electron interaction takes the form of a \emph{dipole-dipole} interaction, which is $V(r) = e^2/r - e^2/\sqrt{r^2+(2d)^2}$ and at distances $r \gg d$ becomes $V(r) \sim e^2 d^2/r^3$.

Now remember that the entire argument for the existence of the WC state was that at large inter-electron distance $r$, the Coulomb interaction $E_C \propto 1/r$ beats the Fermi energy $E_F \propto 1/r^2$. But this is no longer correct if the Coulomb interaction becomes $\propto 1/r^3$. Consequently, when the gate distance is finite, the WC must melt again at low enough densities due to the screening effect \cite{spivak2004phases, skinner2010simple, Valenti_2025}. Consequently the WC-FL phase transition is \emph{re-entrant}, as depicted in Fig.~dipoles(b). So far no experiments have observed the ``dipole-like'' WC-to-FL transition, since it tends to be associated with such low density that disorder or finite temperature can easily overwhelm it.

\begin{figure}[tb!]
	\centering
	\includegraphics [width = 0.9\textwidth]{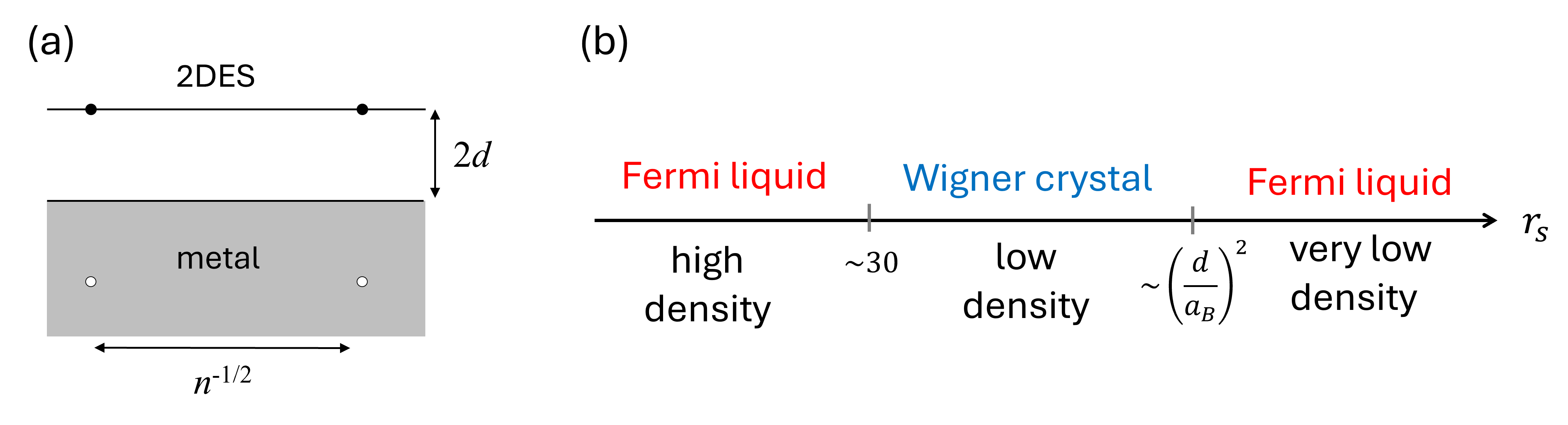}
	\caption{(a) When the distance $n^{-1/2}$ between neighboring electrons (black points) is large compared to the distance $d$ from the metal gate, electrons are essentially dipoles due to the screening effect of their image charges (white points) in the gate. Their interaction $V(r) \sim e^2 d^2/r^3$. (b) This screening effect modifies the phase diagram, producing melting of the WC at very low densities. Notice that if $d/a_B$ is not too large, then there is no WC phase at all.} 
	\label{fig:dipoles}
\end{figure}

\subsubsection{Aside: proof that something like microemulsions must exist}
\label{sec:Kivelsonproof}

\textit{[This subsection is more or less a verbatim repetition of an argument given to me by Steve Kivelson over lunch using a few napkins and a whiteboard marker. These ideas are basically present in Refs.~\cite{spivak2003phase, spivak2004phases, jamei2005universal}, but I liked the napkins + whiteboard marker version better. I also wrote more or less this same text in Ref.~\cite{joy2023upper}.]}

The goal of this section is to argue that, despite the large energy penalty associated with making big domains with different density, it is not possible for an electron system to proceed directly from one uniform phase to another if there is a discontinuity in chemical potential. In other words, direct first-order phase transitions without phase coexistence are not allowed.

The argument proceeds by considering the system at the critical density $n = n_c$ for which the two phases have equal energy (the nominal location of the first-order transition) and showing that one can construct a trial microemulsion state with the same average density but lower free energy than either pure phase. The existence of such a lower-energy state implies that the system must not pass directly from a pure phase 1 to a pure phase 2 as a function of increasing average density, and a direct first-order transition between the two pure phases is precluded. The argument is simplest if one assumes that the interaction between electrons is of the form $V(r)=k/r^\alpha$, where $1<\alpha<2$. The limit $\alpha \rightarrow 1$ is discussed at the end of this section. 

Consider a trial state consisting of alternating stripes of phases 1 and 2, each with the same width $\ell$ (assumed to be much longer than the inter-electron spacing) and having uniform electron density $n_1 = n_c - \delta n$ (phase 1) and $n_2 = n_c + \delta n$ (phase 2), so that the global average density is $n_c$. The free energy per area of this phase, relative to either uniform phase at $n=n_c$, can be written
\be
\delta f = f_{\text{bulk}} + f_{\text{Coulomb}} + f_{\text{surface}}.
\ee

The first contribution, $f_{\text{bulk}}$, represents the change in free energy (per unit area) associated with shifting some electrons from phase 1 to phase 2, which has a lower chemical potential. To leading order in $\delta n$, $f_{\text{bulk}}$ is given by:
\be
f_{\text{bulk}} = -\frac{\delta n}{2} (\mu_1(n_c) - \mu_2(n_c)).
\ee

The second contribution, $f_{\text{Coulomb}}$, corresponds to the electrostatic energy cost associated with the two stripe regions having an overall net charge $\pm e \delta n$ per unit area:
\be
f_{\text{Coulomb}} = A(e\delta n)^2V(\ell)\ell^2,
\ee
where $A$ is a numerical constant of order unity. 

The final contribution, $f_{\text{surface}}$, represents the energy required to create an interface between the two phases:
\be
f_{\text{surface}} = \frac{\sigma}{\ell},
\ee
where $\sigma$ is the interfacial tension.

Keeping $\ell$ fixed, we minimize $\delta f$ with respect to $\delta n$, which gives
\be
\delta n = \frac{\mu_1(n_c) - \mu_2(n_c)}{4Ake^2\ell^{2-\alpha}}.
\ee
The corresponding (minimum) free energy change is given by:
\be
\delta f = -\frac{\left(\mu_1(n_c) - \mu_2(n_c)\right)^2}{16Ake^2\ell^{2-\alpha}} + \frac{\sigma}{\ell}.
\label{eq: deltafSK}
\ee
Notice that so long as $1<\alpha<2$, the change in free energy $\delta f$ is negative whenever $\ell$ is sufficiently long. Thus, we have constructed a trial microemulsion state that has a lower energy than either uniform state, and there must not be a direct first-order transition between two uniform states.

In the limit $\alpha \rightarrow 1$, both terms in Eq.~(\ref{eq: deltafSK}) are proportional to $1/\ell$, and the analysis is inconclusive. However, a more carefully constructed trial state, in which the electron density is allowed to vary within each stripe, still yields a negative value of $\delta f$ in the limit of very long $\ell$ \cite{jamei2005universal}.

\subsection{How can I tell that I have a Wigner crystal?}
\label{sec:WCsignatures}

So you think you've made a WC. How do you prove it? What are the experimental signatures of Wigner crystallization?

The obvious answer is that you take a picture of the WC using some scanning technique like STM. This was done in Refs.~\cite{tsui_direct_2024, xiang2025imaging} (and these studies produced the very first direct images of the WC, 90 years after the idea was first proposed), but these experiments are difficult and generally give you a picture only of a limited region of space.

Another idea is to find some other excitation that travels through the material, ``on top" of the WC electrons, and to look for this excitation to be affected by the Wigner lattice. Such an excitation (assuming it responds to the electrons) sees a periodic potential with periodicity given by the Wigner lattice. Consequently, the excitation lives in a Brillouin zone defined by the electron system, and it can experience zone folding and umklapp scattering by the WC phonon modes. This method was used in Ref.~\cite{smolenski_signatures_2021}, for example, which found that excitons (electron-hole bound states) exhibited umklapp scattering from the Wigner lattice that could be continuously controlled by gating the electron system to different densities. 

Of course, the most natural and straightforward thing to measure about an electron system is its electrical resistance. You might think that the WC is a metal because there are no energy gaps: there is no ``filled band'' or even a ``filled lattice" -- if we try to add another electron then the lattice constant of the Wigner lattice simply adjusts continuously to admit the additional electron. But the electrons are not exactly free, in the sense that they confine each other. So if the WC is to conduct electricity, it can only do so via a collective sliding mode where all the electrons slide together. This mode is gapless if there is literally zero disorder (it is essentially a $q=0$ longitudinal phonon mode). But in the presence of any disorder at all, the sliding mode gets ``pinned''. That is, if there are any potential wells in the system created by disorder, then an electron from the WC can become trapped in that potential well. Once a single electron is trapped, the whole WC is inhibited from sliding. So no current is produced by a weak electric field. Instead, the current only begins to flow when a sufficiently strong electric field is applied that the electric potential landscape becomes ``tilted'' to the point that the previous potential well no longer has a minimum. The consequence is that the $I$-$V$ curve of a Wigner crystal exhibits ``pinning'' behavior, in which no current flows until a critical voltage is applied:

\begin{figure}[h!]
	\centering
	\includegraphics [width = 0.8\textwidth]{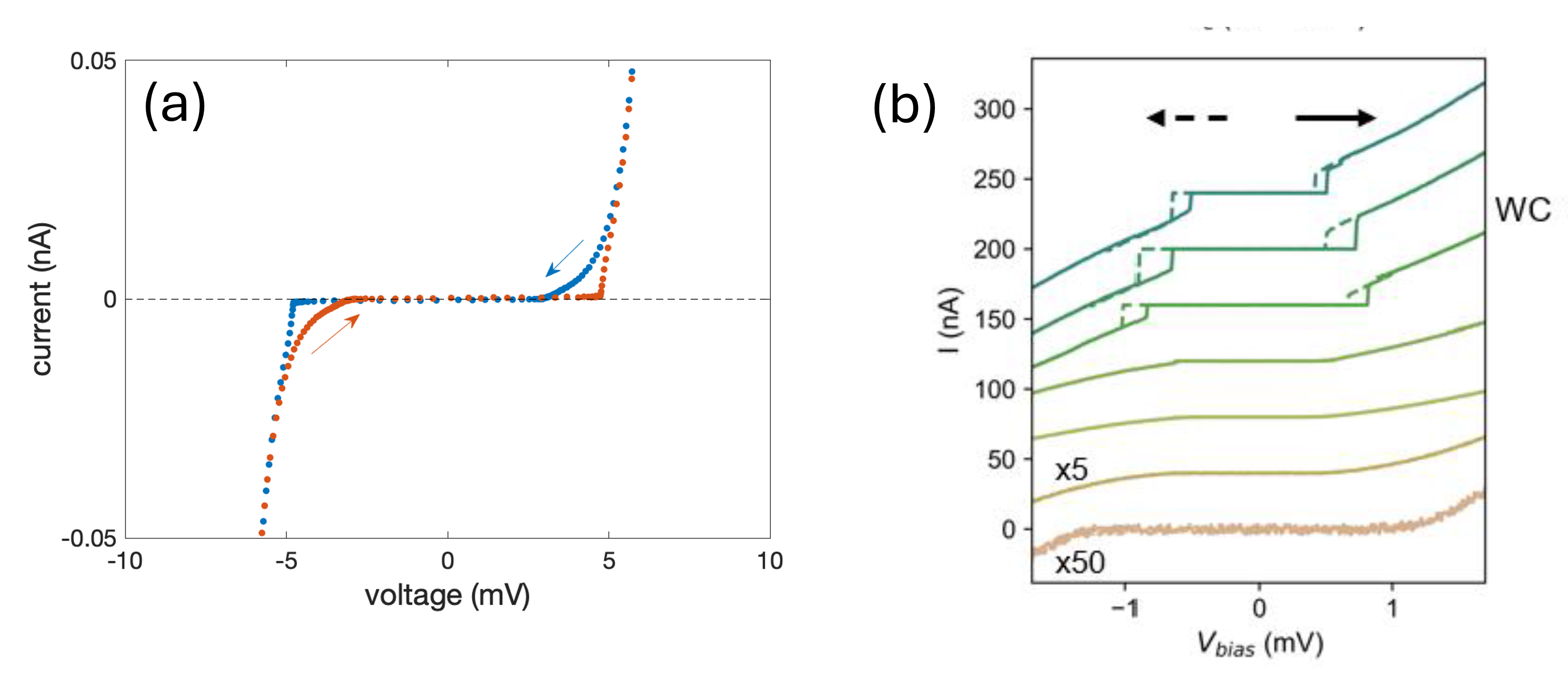}
	\caption{``Pinning" behavior in the $I$-$V$ curves of a 2DES in the WC regime, shown for experiments in (a) ZnO \cite{falson_competing_2022} and (b) rhombohedral six-layer graphene \cite{han2026evidence}.} 
	\label{fig:pinning}
\end{figure}

\noindent The $I$-$V$ curves of a WC are often hysteretic (i.e., once the current is flowing it persists to a smaller value of the electric field), for reasons that are not entirely understood.

In principle, a WC at finite temperature does have a nonzero Ohmic conductivity even when it is pinned. This Ohmic conductivity is associated with a small concentration of freely mobile ``defects'' in the WC, namely negatively-charged interstitial defects and positively-charged vacancies \cite{shklovskii_coulomb_2004}. These defects are discussed in more detail in Sec.~\ref{sec:self-doping}.

WCs are also unusual in terms of their thermodynamic properties, due to the negative sign in Eq.~(\ref{eq:Eclassical}). That is, to leading order, the energy per unit area of a WC is $f \sim - e^2 n^{3/2}$, which means that the chemical potential $\mu = df/dn$ is also negative, and so is the ``inverse compressibility'' $d\mu/dn \sim - e^2/n^{1/2}$. Negative compressibility is a seemingly strange idea; usually, having a negative value of $d\mu/dn$ implies an instability -- if a charge-neutral substance were to have negative $d\mu/dn$, it would imply that the system could lower its energy by spontaneously coalescing densely into some parts of the volume and leaving other parts empty. For the 2DES, fixing the jellium charge prevents this instability, and indeed the compressibility is negative only if one imagines that changing $n$ increases the jellium density simultaneously with the electron density. (Equivalently, one could say that the ``negative compressibility'' implies that electrons apply a compressive stress on the jellium background, with each electron trying to pull its Wigner cell of neutralizing background charge closer to itself.)

Nonetheless, the negative value of $d\mu/dn$ for a WC is directly observable in experiments \cite{eisenstein1992negative, eisenstein1994compressibility}. The trick is to measure the differential capacitance $dQ/dV$ of the capacitor setup shown in Fig.~\ref{fig:capacitor}. The negative energy $E \sim - e^2 n^{1/2}$ of the WC implies that the energy of the capacitor is somewhat lower than the usual value $U = Q^2/(2C)$, where $C$ is the usual ``geometric capacitance'' associated with a parallel plate capacitor. Consequently, the capacitance appears to be enhanced over the geometric value by an amount that is proportional to $- d\mu/dn \propto e^2/n^{1/2}$ \cite{bello_density_1981}. In other words, as the density of the WC is lowered, the capacitance becomes increasingly enhanced. This capacitance enhancement by negative compressibility is a key qualitative feature of Wigner crystals. While it isn't clear that \emph{only} WCs have this feature (one can imagine a correlated liquid where electrons do a pretty good job of avoiding each other, and therefore have a negative energy overall), it is a feature that all WCs must have.

\subsection{Spin ordering of the Wigner crystal}
\label{sec:spinordering}

So far we have not said anything about the spin ordering of the WC, and indeed our semiclassical ``Einstein phonon" picture of the WC is incapable of making any statements about the electron spin. So what is the nature of spin ordering in the WC? This is a surprisingly interesting and intricate question, both in the jellium model and beyond.

First off, as usual in electron systems, the spin ordering in a 2DES has nothing to do with the fact that the electron is itself a little magnet. I mean that the spin magnetic moment of an electron of course produces a magnetic field around the electron, and other electrons can experience this magnetic field, but the associated magnetic forces are puny compared to the electrostatic forces between electrons. Instead, ``magnetic'' interactions arise from the exchange interaction, which is another way of saying that electrons choose whether or not to align their spin by considering only the tradeoff between quantum kinetic energy and electrostatic energy. In a sense, electrons use the Pauli exclusion principle as a loophole which they may or may not want to exploit \cite{brian_skinner_where_2015}: if they align their spins, then they are guaranteed by Pauli exclusion to avoid each other spatially,\footnote{An under-utilized variation of the Pauli exclusion principle is ``two electrons with the same spin cannot be in the same place at the same time.''} %
and thus their Coulomb energy is lowered. But this spin alignment also forces the electrons into higher kinetic energy states. This tradeoff is the basic nature of the exchange interaction and the origin of magnetism.

When most of us learn about the exchange interaction, it is in the context of something like a two-electron problem with interactions, such as a two-site Hubbard model. In this context, the exchange interaction generically produces antiferromagnetism, or in other words a term in the spin Hamiltonian that looks like $+J \vec{S}_1 \cdot \vec{S}_2$, where $\vec{S}_1$ and $\vec{S}_2$ are the two interacting spins and $J \sim t^2/U$ is the resulting coupling that comes from the exchange ($t$ is the hopping energy between neighboring sites and $U$ is the on-site interaction energy). The interaction is antiferromagnetic because if the two electron spins are antialigned then their associated wave functions are able to ``spread out'' a bit onto both sites (e.g., hopping from one site to another is not forbidden by Pauli exclusion), thereby lowering the quantum mechanical energy. This spreading out comes at the cost of an interaction energy, which appears as the denominator $U$ in second-order perturbation theory \cite{girvin2019modern}. It is natural to expect that the WC would have an antiferromagnetic interaction for similar reasons. And indeed, if one considers only the interaction between pairs of neighboring electrons, the interaction \emph{is} antiferromagnetic.

But in a full WC, there are many exchange processes that can happen, beyond just a nearest-neighbor exchange, and in principle all of these processes must be taken into consideration. We can consider, for example, exchanging the positions of three nearest-neighboring electrons simultaneously by shuffling them around either clockwise or counterclockwise: this process favors ferromagnetism. Or we can consider exchanging four electrons in a rhombus of nearest neighbors, or we can consider five-electron exchanges, and so on. The most general form of the effective Hamiltonian for the spin degree of freedom looks like \cite{kim_interstitial-induced_2022, Kim2024dynamical}
\be 
H_\textrm{spin} = \sum_{a}(-1)^{n_a} J_a \left(\mathcal{P}_a + \mathcal{P}_a^{-1} \right),
\label{eq:Hspin}
\ee 
where $a$ labels a ring exchange process involving $n_a$ electrons, $J_a > 0$ is the exchange constant for such processes, and $\mathcal{P}_{a}$ is the permutation operator associated with the exchange process $a$.\footnote{A two-spin permutation operator  $\mathcal{P}_{ij}$ exchanges the spin states between electrons $i$ and $j$. It can be written in terms of the spin operators $\vec{S}_i, \vec{S}_j$ of the two spins as $\mathcal{P}_{ij} = (1 + 4 \vec{S}_i \cdot \vec{S}_j)/2$. For a three-spin permutation process, one can write $\mathcal{P}_{ijk} = \mathcal{P}_{jk}\mathcal{P}_{ij}$ and its inverse $\mathcal{P}_{ijk}^{-1} = \mathcal{P}_{ij}\mathcal{P}_{jk}$, and in general one can construct any permutation operator $\mathcal{P}_a$ by imagining a chain of two-spin permutations, proceeding either clockwise or counterclockwise. }  So exchange processes $a$ with an even number of electrons $n_a$ have a positive coefficient and favor antiferromagnetism, while exchange processes with an odd number of electrons have a negative coefficient and favor ferromagnetism.

\begin{figure}[tb]
	\centering
	\includegraphics [width = 0.6\textwidth]{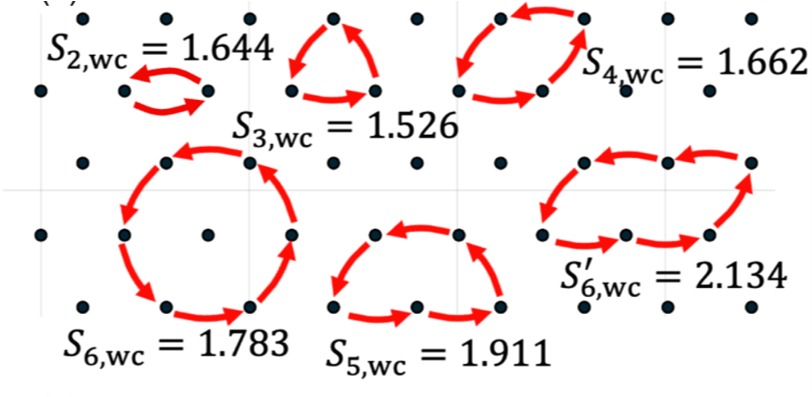}
	\caption{A few semiclassical exchange paths for the WC, taken from Ref.~\cite{kim_interstitial-induced_2022}. The number listed is the (imaginary) action associated with the exchange in units of $\hbar \sqrt{r_s}$.} 
	\label{fig:exchange_paths}
\end{figure}

The existence of all the extra exchange terms $n_a > 2$ in Eq.~(\ref{eq:Hspin}) usually doesn't matter in the strongly-interacting regime, since these terms are usually increasingly suppressed when more electron are involved in the exchange. For example, in the Hubbard model, a three-electron ring exchange process has an amplitude $J_3 \sim t^3/U^2$, a four electron exchange process has an amplitude $J_4 \sim t^4/U^3$, and so on, so that if $U \gg t$ these processes become increasingly unimportant as $n_a$ increases. Consequently, one usually keeps only the lowest-order, two-electron exchange process $J_2 \sim t^2/U$. You can also think about the exchange processes in a path integral way: imagine a semiclassical trajectory $P$ by which some number of electrons physically trade places. This process has some (imaginary) action $S$ associated with it: $S = \int_P \vec{p} \cdot d\vec{s}$, where $\vec{p}$ is the (imaginary) momentum associated with passing under a potential energy barrier along the $\vec{s}$ coordinate. One should in general sum over all possible paths $P$, but deep in the WC limit all such paths involve a large action $|S| \gg \hbar$, and the exchange constant is dominated by the path of least action, $J \propto \exp[i S_\textrm{min}/\hbar]$, where $S_\textrm{min}$ is the action associated with the least-action path. Such a path will generally involve the electrons passing over a saddle point of the potential energy landscape. Since an $n_a$-electron exchange process involves moving $n_a$ electrons simultaneously, it is natural to expect that the action increases with $n_a$, and therefore the amplitude $J_a$ decreases exponentially with increasing $n_a$. 

But there is a surprising oddity about the WC: the three-electron exchange has a smaller least-action path than the two-electron exchange does: $S_3 < S_2$. There is no obvious way to understand this inequality; $S_3$ is only smaller than $S_2$ by a numerical factor.\footnote{Specifically, $S_3 \simeq 1.53 \hbar \sqrt{r_s}$ and $S_2 \simeq 1.64 \hbar \sqrt{r_s}$ \cite{Kim2024dynamical}.} But the basic intuition is that in a triangular lattice, simultaneously rotating three electrons is unusually ``soft'', so that the increase in action associated with having to move three electrons rather than two is more than compensated by having a lower potential barrier to pass through. This is the kind of surprising result that can only appear by thinking about electrons moving through a smooth potential in real space, rather than using a tight-binding picture with discrete electron ``hops''.

Since the three-electron exchange term favors ferromagnetism, the dominance of $J_3$ over $J_2$ implies \emph{ferromagnetic} spin ordering deep in the WC state. Quantum Monte Carlo studies \cite{drummond2009phase, azadi2024quantum} (and some experimental evidence \cite{zhang2025transportevidencewignercrystals, falson_competing_2022}) suggest a window of antiferromagnetism very close to the WC-FL transition (where multiple different exchange processes are of similar magnitude), but deep in the WC the ordering is ferromagnetic.\footnote{Of course, \emph{all} exchange processes become exponentially weak deep in the WC phase, owing to the very small overlap between electron wave functions. So in any given experiment it will probably be the case that deep in the WC phase one has a classical paramagnet, since $k_B T$ becomes larger than any $J$.}

\subsection{Self-doping instability}
\label{sec:self-doping}

This old-fashioned physics about virtual processes has produced a surprising new theoretical proposal about the jellium model WC \cite{Kim2024dynamical}. The proposal is that the WC can become unstable with respect to spontaneous formation of defects in the Wigner lattice, and that these defects are mobile even when the WC itself is not. In this scenario, the WC becomes a ``metallic electron crystal'', which conducts electricity like a metal even when the parent WC lattice is pinned and insulating. 

The relevant types of defects for this proposal are vacancies and interstitials in the Wigner lattice. A vacancy is formed by removing an electron from one of the Wigner lattice sites, and an interstitial is formed by adding an extra electron to the center of one of the triangles of the Wigner lattice. After allowing the electrons in the vicinity of the defect to relax, the defects look something like this:

\begin{figure}[htb]
	\centering
	\includegraphics [width = 0.8\textwidth]{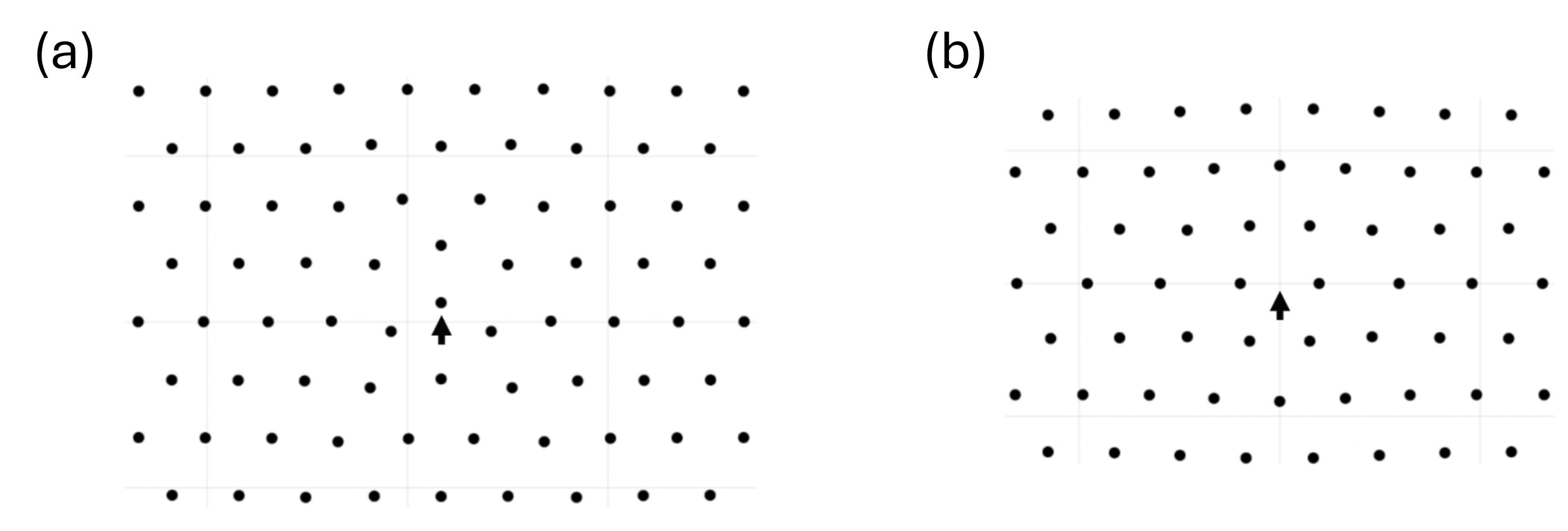}
	\caption{An interstitial defect (a) and a vacancy (b) in the Wigner lattice. Taken from \cite{Kim2024dynamical}.} 
\end{figure}

\noindent Creating such a (localized) defect is associated with a finite electrostatic energy $E_0 \sim e^2 n^{1/2}$, which is positive.

For a classical system at zero temperature, these defects are fixed in place. But quantum mechanically, the defects are allowed to move by a virtual process that is something like the ones discussed in Sec.~\ref{sec:spinordering}. That is, there is a finite (imaginary) action $S$ associated with a process that involves rearranging the positions of a bunch of neighboring electrons in such a way that the vacancy or interstitial is shifted by one lattice site. Consequently, one can say that the defect has a finite hopping amplitude $t \propto \exp[i S/\hbar]$ for hopping between neighboring sites of the Wigner lattice. In this sense, a single defect resides in a band with width $6 t$.\footnote{$6t$ is the bandwidth for a noninteracting tight-binding model of hopping on either a triangular lattice (where vacancies reside) or a honeycomb lattice (where interstitials reside).} A state with zero momentum (e.g., a uniform superposition of all possible locations) has an energy $-3 t$ relative to the band center.

The idea of the ``self-doping instability" is that a (delocalized) defect state at the bottom of the band has energy $E_0  - 3t$, and there might be some range of $r_s$ for which this energy is negative. If this happens, then the system can lower its energy by creating defects, and these defects can be mobile (albeit heavy, presumably). If the defects appear with some finite concentration, then one has a ``metallic electron crystal''. The calculations of Ref.~\cite{Kim2024dynamical} estimate that this scenario can appear in the jellium model WC when $r_s \lesssim 70$, and in this range interstitials have negative energy. (The range of $r_s$ for which creating an interstital-vacancy pair from the pristine WC is energetically favorable is estimated as $r_s \lesssim 45$). 

This idea has become especially interesting in light of recent experiments \cite{han2026evidence} in 5- and 6-layer rhombohedral graphene (discussed in Sec.~\ref{sec:materials}). These experiments observe something that looks very much like a WC, with a pinning-like $I$-$V$ curve and negative compressibility. However, by slightly modulating some parameters (the values of two gate voltages), they find an abrupt transition to a metallic state with opposite sign of the Hall conductivity (e.g., opposite sign of the carriers). This state is interpreted as a ``metallic electron crystal'' with mobile vacancies, similar to what is proposed in Ref.~\cite{Kim2024dynamical} (and seemingly supported by additional theory calculations in Ref.~\cite{dong2026crystals}), although the result is still quite new and there is much that remains to be understood.

\section{Newfangled physics: Wigner crystals with Berry curvature}

\subsection{Newfangled materials}
\label{sec:materials}

As explained in Sec.~\ref{sec:whatis}, the WC state appears in the regime where all relevant energy scales are small due to the low electron density. Thus, if one hopes to realize WC physics, the material in which it is realized needs to be \emph{clean} (so that the disorder energy scale is small) and \emph{cold} (so that $k_B T$ is small). The traditional champions of ``clean and cold'' for 2D electron physics are silicon (most notably, inversion layers at the surface of silicon) and gallium arsenide (specifically, quantum wells made from GaAs-AlGaAs heterostructures). Many decades of effort into growing and purifying these materials have enabled them to have spectacularly low defect concentrations. But unfortunately both of them have relatively high dielectric constant ($\varepsilon \approx 12$ for Si, $\varepsilon \approx 13$ for GaAs) and relatively low effective mass ($m \approx 0.19m_0$ for Si inversion layers, $m \approx 0.07m_0$ for GaAs, where $m_0$ is the bare electron mass), and this has the effect of pushing the WC to very low density (and, consequently, very low temperature). 

Among traditional semiconductor materials, zinc oxide (ZnO, perhaps most widely known as the active ingredient in many sunscreens) has emerged as a strong contender for realizing WC physics, following years of work into making 2D heterostructures \cite{falson2018review}. While its compositional purity is probably still not as high as Si or GaAs, ZnO has the advantage of having a higher effective mass ($m \approx 0.3m_0$) and lower dielectric constant ($\varepsilon \approx 8.5$). These numeric differences might not seem very significant, but since $r_s \propto 1/\sqrt{n (\varepsilon/m)^2}$, they mean that the WC state appears at a $> 40 \times$ higher density in ZnO than in GaAs, and therefore the WC state is easier to study. Relatively clear features of the FL-WC transition have been observed in ZnO \cite{falson_competing_2022}.

Perhaps even more noteworthy has been the progress on the two-dimensional semiconducting materials known as the transition metal dichalcogenides (TMDs). These are two-dimensional semiconductors with relatively large effective mass (ranging from $0.3 m_0$ to $0.8 m_0$), and since they are two-dimensional, the effective dielectric constant is determined by the substrate, which can be relatively small (as low as $\varepsilon \approx 3$ for boron nitride, for example). The TMDs can also be stacked, layered, and twisted in different ways to modify the band structure. There are a number of recent experiments reporting WC physics in this context (e.g., Refs.~\cite{zhang2025transportevidencewignercrystals, xiang2025imaging, zhou_bilayer_2021, sung2025electronic, smolenski_signatures_2021}). 

But, in my opinion, the most dramatic ``new'' WC physics is appearing in multilayer graphene, and most particularly in ``rhombohedral'' (ABC-stacked) graphene. These materials are gapless in their native state, meaning that the conduction and valence bands meet at two points in momentum space (the K and K' points).  It turns out that a gapless material cannot host a WC due to the very large interband dielectric response associated with creating (virtual) electron-hole pairs. This dielectric response kills the long-ranged part of the Coulomb repulsion and prevents WCs from forming \cite{joy2023wigner}. But application of a perpendicular electric field opens a band gap that simultaneously kills the interband dielectric response and can create a very flat band that is especially propitious for the formation of WCs. There are numerous recent reports of WC-type states in rhombohedral graphene \cite{han2026evidence, lu2024fractional, han_signatures_2025} (alongside other wild things like chiral superconductivity and integer and fractional Chern insulating states at zero magnetic field; these may be more exciting but don't get distracted because we're talking about Wigner crystals right now).

But in addition to having a controllable band structure (which in many cases can be qualitatively different from the usual parabolic bands with well-defined effective mass $m$), rhombohedral graphene offers the qualitatively new ingredient of \emph{Berry curvature}. The question ``how does Berry curvature alter the nature of the WC state?" is still very much under investigation, but in the remainder of these lecture notes I'll try to briefly mention a few recent ideas.

\begin{figure}[htb]
	\centering
	\includegraphics [width = 0.8\textwidth]{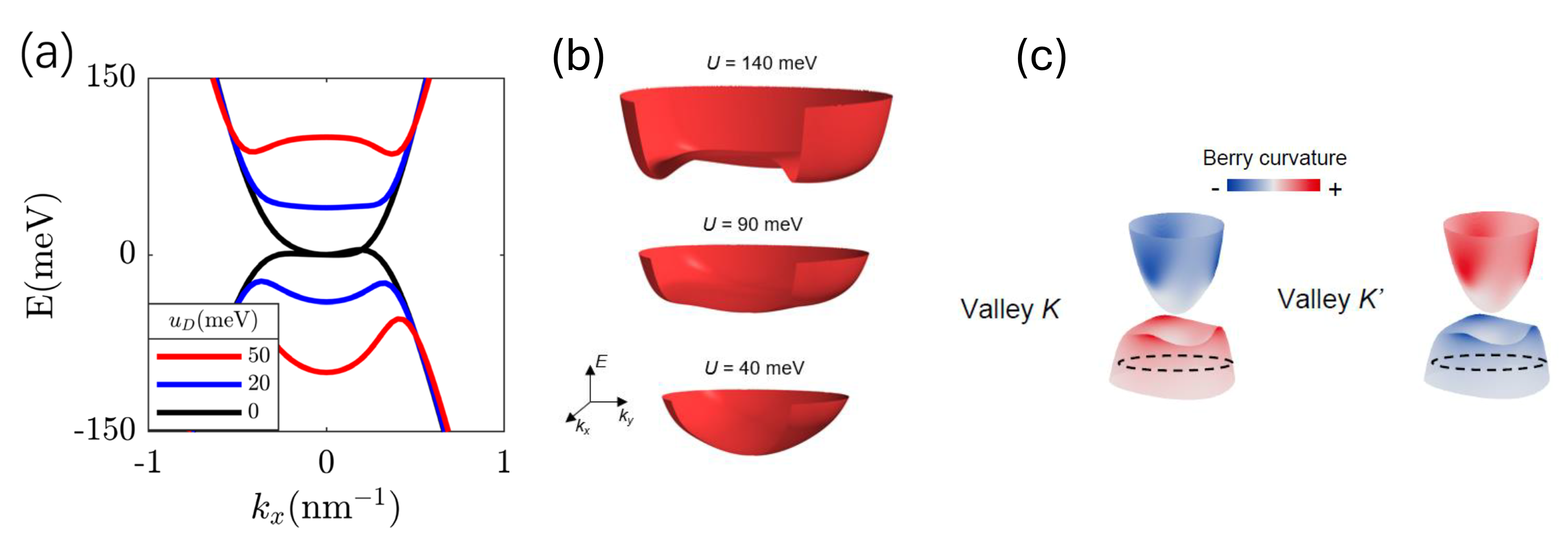}
	\caption{(a) The conduction and valence bands in rhombohedral 5-layer graphene for different values of the interlayer potential (from Ref.~\cite{tan2024parent}). (b) The shape of the conduction band in rhombohedral 6-layer graphene for different value sof the interlayer potential (from Ref.~\cite{han2026evidence}). (c) The distribution of Berry curvature among the two bands in the vicinity of the K and K' points in 5-layer rhombohedral graphene (from Ref.~\cite{han_orbital_2023}).} 
	\label{fig:dispersions}
\end{figure}

\subsubsection{Aside: What is Berry curvature?}

No one talked about Berry curvature when I was a graduate student in the suddenly-paleolithic years of 2007-2011, but by now it has been the subject of many review articles and textbook chapters.\footnote{Girvin and Yang's textbook is probably the best \cite{girvin2019modern}, but I also really like this recent review \cite{Verma_2026_quantum}. You can find my own attempt at a conceptual introduction in Ref.~\cite{ramirez2020dawn}.} Nonetheless, it is probably the friendly thing to do to give you a lightning review of what Berry curvature is and how one can think about it conceptually. So here's how I personally like to think about Berry curvature:

Imagine if there was a single electron described by a wave function $\psi(\vec{r})$ and I asked the question ``where is the electron?". You could rightly reply ``That's not really a well-defined question. Unless the wave function happens to be a position eigenstate (i.e., a delta function of $\vec{r}$), it resides in a superposition of many positions. But if you insist on me giving you a single number then I guess you could calculate the expectation value of the position operator:
\be 
\langle \vec{r} \rangle = \langle \psi | \vec{r} | \psi \rangle = \int d^dr \, \vec{r} \left| \psi(r) \right|^2. "
\ee 
And if, for whatever reason, I gave you the wave function $\tilde{\psi}(k)$ written in momentum space, you could just as well calculate the expectation value of position using the expression for the position operator in $k$-space, $\vec{r} = i \vec{\nabla}_k$, so that
\be 
\langle \vec{r} \rangle = i \langle \tilde{\psi} | \vec{\nabla}_k | \tilde{\psi} \rangle = i \int \frac{d^dk}{(2\pi)^d}  \, \tilde{\psi}^*(k) \vec{\nabla}_k  \tilde{\psi}(k).
\ee 

Now suppose that I gave you a wave function that is a momentum eigenstate: $\psi_k(\vec{r}) = u_k(\vec{r}) e^{i \vec{k} \cdot \vec{r}}$  (where $u_k(\vec{r})$ is the Bloch function associated with the crystal lattice in which the electron resides\footnote{When I was in grad school, the purpose of teaching Bloch's theorem was seemingly to say ``Look, I know that momentum eigenstates in a crystal are not \emph{literally} plane waves; they are actually plane waves modulated by a periodic function. But the Bloch functions are not actually important so let us never speak of them again.'' In this sense I often think of the 21st-century fascination with quantum geometry as ``the revenge of the Bloch functions.''}) and I asked you ``where is the electron?". This might seem like an exceptionally dumb question, since a momentum eigenstate is extended over the entire system and is not really ``located'' anywhere. Nonetheless, you could still calculate $\langle \vec{r} \rangle$, and it would give you a value that only depends on the Bloch functions:
\be 
\langle \vec{r} \rangle \equiv \vec{A}(\vec{k}) = i \langle u_{\vec{k}} | \vec{\nabla}_k | u_{\vec{k}} \rangle.
\ee 
The quantity $\vec{A}(\vec{k})$ is a vector field with units of position: it tells you ``the center-of-mass location of the wave function" (whatever that might mean) as a function of its momentum $\vec{k}$. Of course, there is a huge caveat here, which is that $u_{\vec{k}}$ is not uniquely defined. In fact, $u_{\vec{k}}$  can be multiplied by a factor $e^{i \theta_k}$, where $\theta_k$ is any smooth function of $\vec{k}$, without changing the physical electron probability density $|\psi_k(r)|^2$. Different choices for the function $\theta_k$ (``gauge choices'') produce different answers for $\vec{A}(\vec{k})$ (that is, the quantity $\vec{A}(\vec{k})$ is ``not gauge invariant''). So we can't take the meaning of $\vec{A}(\vec{k})$ too literally: it is not really true that an electron with momentum $\vec{k}$ is actually ``centered at position $\vec{A}(\vec{k})$''. But there \emph{is} a real meaning to how $\vec{A}(\vec{k})$ changes as you change the momentum $\vec{k}$, and these changes are associated with a physical electron current.

Specifically, imagine taking an electron with a given momentum $\vec{k}$ on an excursion through momentum space, accelerating and decelerating it in some way via time-changing electric fields. Over the course of the excursion, the wave function acquires a ``geometric'' phase $\varphi = \int \vec{A}(\vec{k}) \cdot d\vec{k}$, in addition to the usual dynamical phase $\frac{1}{\hbar} \int E(t) dt$. (You can think about the phase $\varphi$  as something like the usual phase accumulated by a wave, $\int \vec{k} \cdot d\vec{x}$, which is just $2 \pi \times (\textrm{\# of wavelengths in the path})$, but we have written it in terms of a path in momentum space rather than physical space.) The phase $\varphi$ is observable as a \emph{Berry phase} if we imagine the excursion to be along a closed path $P$ in momentum space (i.e., returning to its original momentum) -- such a phase describes constructive or destructive interference, either of an electron with itself or of two different electrons meeting after following two parallel arms of the path $P$. So $\varphi$ is a real, observable, gauge-invariant quantity. For closed paths we can use Stokes' theorem
\be 
\varphi = \oint_P \vec{A}(\vec{k}) \cdot d\vec{k} = \iint_S (\vec{\nabla}_k \times \vec{A}) \cdot d\vec{S},
\label{eq:Berryphase}
\ee 
where $S$ is a surface (in momentum space) enclosed by the path $P$. Since $\varphi$ is a real, observable quantity, the quantity $\vec{\nabla}_k \times \vec{A}$ must also be a real, observable quantity (i.e., it \emph{is} gauge invariant) -- we call it the ``Berry curvature'' $\vec{\Omega}(\vec{k})$.

Notice that Eq.~(\ref{eq:Berryphase}) for the Berry phase looks like the equation for the Aharonov-Bohm phase acquired by an electron in a magnetic field:
\be 
\varphi_{\textrm{AB}} = -\frac{e}{\hbar} \oint_P \vec{A}_{\textrm{B}}(\vec{r}) \cdot d\vec{r} = -\frac{e}{\hbar} \iint_S \vec{B} \cdot d\vec{S},
\ee 
where now $P$ is a path in real space, $\vec{A}_{\textrm{B}}$ is the (not gauge invariant) magnetic vector potential, and $\vec{B} = \vec{\nabla} \times \vec{A}_{\textrm{B}}$ is the (gauge invariant) magnetic field.

In this sense we often say that Berry curvature $\vec{\Omega}(\vec{k})$ is like a ``magnetic field in $k$-space''. Its value depends not on the electron position, but on its momentum. Physically, Berry curvature arises because electrons live in a superposition of different orbitals (e.g., on different atoms) within the unit cell of the crystal. As the momentum of the electron shifts, the weight of this superposition can shift from one orbital to another, and this shift can produce a nontrivial phase and a physical charge current.\footnote{The ``anomalous velocity'' $\vec{v}_A = \dot{\vec{k}}\times \vec{\Omega}$ is an example of a physical current induced by Berry curvature. It arises when the weight of the electron wave function shifts within the unit cell in response to a changing electron momentum (usually induced by an electric field, in which case $\dot{\vec{k}} = -e \vec{E}/\hbar$). The anomalous velocity is the analog of the ``E-cross-B drift'' velocity, $\vec{v}_d = (\vec{E} \times \vec{B})/B^2$, produced by an electric field in the presence of a magnetic field. Another example is the ``shift current'' associated with optical excitation of an electron from one band to another: as the electron is excited from one band to another, its orbital character changes and so its position can shift within the unit cell.}

\subsection{Anomalous Hall crystals}

The integer Chern number $C$ in two dimensions is a property of a filled band. It is usually written as an integral of the Berry curvature over all momentum states in the Brillouin zone,
\be 
C = \frac{1}{2\pi} \int_{\textrm{BZ}} d^2k \, \Omega(\vec{k})
\ee
(the Berry curvature $\Omega$ is essentially a scalar in two dimensions, since it always points in the direction perpendicular to the plane). When $C \neq 0$, there are $|C|$ dissipationless edge states running around the bulk of the system, providing a quantized Hall conductance and zero longitudinal resistance, as in the Quantum hall effect.

At first blush, it seems unnatural to apply the concept of the Chern number to the Wigner crystal, since a Wigner crystal is not a ``filled band'' (in fact, it usually occurs when only a tiny fraction of states near the band edge is filled) and its electrons are basically little harmonic oscillators that occupy something like \emph{position} eigenstates (localized wave packets), rather than momentum eigenstates. But on the other hand, any collection of wave packets can always be written as a superposition of plane wave eigenstates if we want to. And there is a sense of a much smaller, filled ``Brillouin zone'' in the WC problem, since the Wigner lattice creates a periodic potential with a large period given by the electron density itself, and the electrons have arranged themselves precisely to have one electron per period. So one can ask whether this collection of plane waves can have a nonzero Chern number. Or, in other words, can a WC have a nontrivial topological invariant just like the quantum Hall state?

This is a bit of a tricky question, since one must calculate the Berry curvature of the ``plane wave eigenstates'' in a self-consistent way, including the electron-electron interaction in addition to the Berry curvature of the band from which the electron wave packets are made. But the answer seems to be yes, and the resulting state is called an ``anomalous Hall crystal''.

An anomalous Hall crystal is a somewhat confusing thing to think about. To first order one can imagine it as a Wigner crystal in the bulk with a dissipationless edge state curling around its edge. Thus, to any bulk probe, it looks basically like the usual WC: it has negative compressibility and packets of charge density arranged in a triangular lattice. But to electrical resistance measurements it looks dramatically different. While a WC at low temperature is an electrical insulator (large longitudinal resistance $\rho_{xx}$) with no Hall effect ($\rho_{xy} \rightarrow 0$), a Hall crystal has vanishing longitudinal resistance ($\rho_{xx} \rightarrow 0$) and quantized Hall effect ($\rho_{xy} =  (1/C)h/e^2$). 

One can arrive at a cartoon picture of a Hall crystal via the following thought experiment. Imagine first a uniform quantum Hall state (I'll call this a ``QHE liquid''), which has an edge state running around the bulk. Now imagine modulating the electron density (say, in stripes), so that there are alternating stripes of QHE liquid and empty regions. Since QHE liquids have chiral edge states, this striping produces alternating pairs of counterpropagating edge states. The last step in this thought experiment is to allow the nearby pairs of counterpropagating edges to hybridize with each other, in the usual quantum mechanical way, and gap out. Then one arrives at something like a Hall crystal: it has strongly modulated stripes of charge density in the bulk and one leftover edge state running around it. Like so:

\begin{figure}[h!]
	\centering
	\includegraphics [width = 0.4\textwidth]{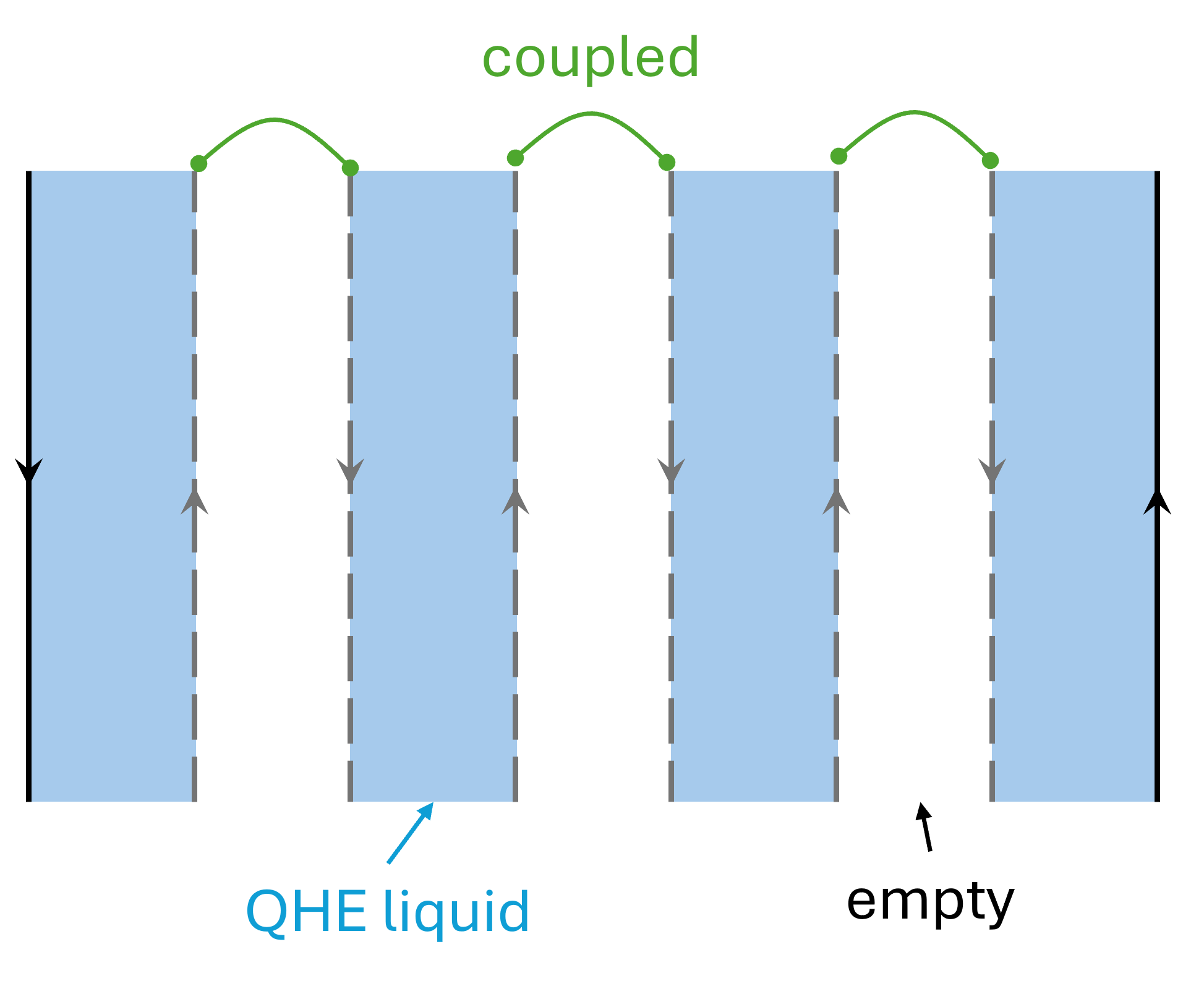}
	\label{fig:coupled_edges}
\end{figure}

\noindent [This thought experiment was explained to me by Ray Ashoori, who said he learned it from Liang Fu. Neither of them is to be held responsible if I have mangled their explanation, \href{https://en.wikipedia.org/wiki/Telephone_game}{telephone}-style.]

The idea of a Hall crystal was first raised in 1989 by Tesanovic and Halperin in the context of the usual quantum Hall effect (where there is magnetic field instead of Berry curvature \cite{tesanovic1989hall}). The idea has been revived in the context of Berry curvature thanks to recent experiments in twisted TMDs and multilayer graphene (e.g., Refs.~\cite{lu2024fractional, su_moire-driven_2025, cai_signatures_2023, park_observation_2023}), which observe integer and fractional quantized Hall conductance in close proximity to a WC state. 

The question of how and when a nonzero Chern number emerges from the old-fashioned WC is a nontrivial theoretical question. One can envision a phase diagram in terms of electron density, temperature, and some parameter that describes the parent electron band (e.g., the Berry curvature near the bottom of the band, or the value of a perpendicular displacement field). I am not going to make any attempt to discuss progress on this question, but here are a few recent theory papers devoted to this question: \cite{tan2024parent, dong_anomalous_2024,  soejima_anomalous_2024, dong2024stability}.

\subsection{WCs with nonzero angular momentum}

It turns out that there is a somewhat simpler kind of transition that can be brought about by Berry curvature, and which is understandable within the ``Einstein phonon" picture of isolated electron wave packets introduced in Sec.~\ref{sec:semiclassics}. The key is to remember that Berry curvature couples to angular momentum in much the same way that magnetic field does, either raising or lowering the energy of states that have nonzero angular momentum, depending on whether the angular momentum is aligned or antialigned with the Berry curvature. If the Berry curvature is strong enough, this coupling can invert the usual ordering of energy levels of the harmonic oscillator, bringing a nonzero angular momentum state to become the lowest energy state. When this happens, the electrons within the WC start rotating in place.\footnote{It is probably important to note that \emph{all} wavepackets have a finite magnetization (often called ``self-rotation'') when they reside in a region of momentum space with nonzero curvature, regardless of whether they have nonzero angular momentum. The jump to finite angular momentum just provides an additional orbital contribution to the magnetization.}

This angular momentum transition becomes simplest to understand in the case where the parent band has a ``Mexican hat" shape, meaning that the band edge forms a ring in momentum space (rather than the usual point in momentum space). The WC electron forms its wave packet by using $k$-components close to this minimum, meaning that in this case $|\tilde{\psi}(k)|^2$ also looks like a ring in momentum space. The flux $\Phi = \int d^2k \, \Omega(k)$ of Berry curvature through the interior of this ring acts something like an Aharonov-Bohm phase. When its value is $|\Phi| > \pi$, the ground state of the WC wave packet has nonzero angular momentum.

\begin{figure}[h!]
	\centering
	\includegraphics [width = 0.5\textwidth]{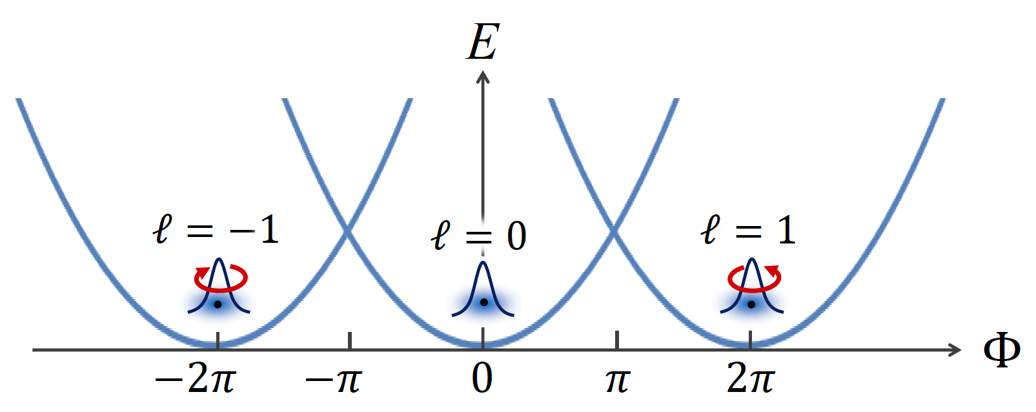}
	\caption{The energy $E$ of an electron wave packet as a function of the Berry flux $\Phi$ through its interior for different values of the angular momentum $\ell$ (figure from Ref.~\cite{joy2025chiral}).}
	\label{fig:BerryFluxHO}
\end{figure}

This phenomenon was called an ``$\ell = 1$ WC" in Ref.~\cite{joy2023wigner} and a ``halo WC" in Ref.~\cite{soejima2025lambda}.

\subsection{Spin ordering in the presence of Berry curvature}

Berry curvature is an orbital phenomenon: it describes the structure of momentum eigenstates and has no explicit connection to the physical electron spin. So one might think that the Berry curvature does not play a role in the spin ordering unless the material has spin-orbit coupling (i.e., a dependence of the electron energy on the relative orientation of its spin and momentum vectors). But this turns out not to be the case: Berry curvature \emph{does} play a key role in determining the spin order of the WC even without invoking any other physics.

The key idea is that the spin order of the WC is determined by physical tunneling trajectories, as explained in Sec.~\ref{sec:spinordering}, and these trajectories correspond to paths in $k$-space also. For example, consider the three-spin exchange process that becomes dominant at large $r_s$. In position and momentum space it looks something like this:

\begin{figure}[H]
	\centering
	\includegraphics [width = 0.6\textwidth]{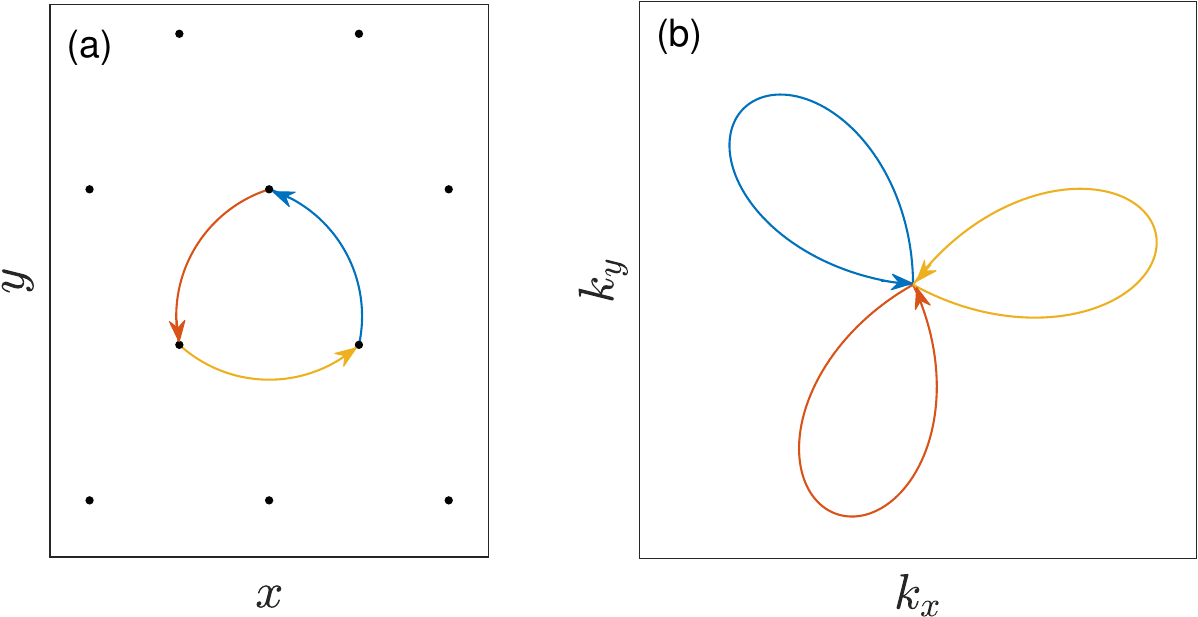}
	\caption{Tunneling trajectories associated with three-electron ring exchange, shown (a) in real space and (b) in $k$-space (from Ref.~\cite{joy2025chiral}).}
	\label{fig:tunneling_paths}
\end{figure}

\noindent Here you should think of the ``momentum'' being plotted as an imaginary quantity,
\be 
k\left(\vec{r}\right) = \sqrt{\frac{2m}{\hbar^2}\left(E - U(\vec{r}) \right)},
\ee
since the electrons are tunneling under a barrier for which the potential energy $U(\vec{r})$ is higher than the total energy $E$. Since the electrons begin and end at a potential minimum, where $U = E$, their momentum-space trajectories similarly begin and return to zero. But along the way the trajectories do some bending, and consequently they encompass an area in $k$-space that is associated with a net Berry phase $\varphi = \int \vec{A} \cdot d\vec{k} = (\textrm{Berry flux inside the $k$-paths})$. Importantly, this phase is opposite in sign for counterclockwise versus clockwise exchange paths.

Consequently, the spin Hamiltonian [Eq.~(\ref{eq:Hspin})] gets modified to include the Berry phases associated with the exchange process. For example, the three-spin exchange term for three neighboring spins $i, j, k$ now looks like
\be 
-J_3 \left( \mathcal{P}_{ijk} e^{i \varphi} + \mathcal{P}_{ijk}^\dagger e^{-i \varphi} \right).
\ee
The permutation operator $\mathcal{P}_{ijk} = \mathcal{P}_{ij} \mathcal{P}_{jk}$ can be written in terms of spin operators as $\mathcal{P}_{ijk} = 1/4 + \vec{S}_i \cdot \vec{S}_j + \vec{S}_j \cdot \vec{S}_k + \vec{S}_i \cdot \vec{S}_k - 2 i \vec{S}_i \cdot (\vec{S}_j \times \vec{S}_k )$.\footnote{You can figure this out yourself, if you want, from $\mathcal{P}_{ij} = (1 + 4 \vec{S}_i \cdot \vec{S}_j)/2$. It's not that fun.} 
In the absence of any Berry phase, that final three-spin term $\vec{S}_i \cdot (\vec{S}_j \times \vec{S}_k)$ vanishes from the Hamiltonian, since it appears with a negative sign in $\mathcal{P}_{ijk}$ and with a positive sign in $\mathcal{P}_{ijk}^{\dagger}$. In this case one is left only with the $\vec{S} \cdot \vec{S}$ terms, producing simple ferromagnetism. But when the Berry phase $\varphi$ is present, the three-spin terms no longer cancel, and the Hamiltonian looks like
\be 
H_{\textrm{spin}} = 
J \sum_{\langle i,j \rangle} \vec{S}_i \cdot \vec{S}_j  + J_\chi \sum_{i,j,k \in \triangle, \triangledown} \vec{S}_i \cdot \left( \vec{S}_j \times \vec{S}_k \right),
\label{eq:Hspinchiral}
\ee 
where $J_\chi = 4 J_3 \sin \varphi$.

This nonzero ``chiral term'' in the Hamiltonian, $ \vec{S}_i \cdot ( \vec{S}_j \times \vec{S}_k )$, is something new and strange. It favors a spin pattern where triangles ($\triangle, \triangledown$) of neighboring spins align themselves at right angles to each other, and it competes with the simple ferromagnetism (or, potentially, antiferromagnetism at not too large $r_s$) associated with the $J$ term in Eq.~(\ref{eq:Hspinchiral}). The existence of this term for WCs has only recently been pointed out \cite{joy2025chiral, kim_exchange_2026} (following similar ideas developed for the Fermi liquid with Berry curvature \cite{Chiral_Stoner_PhysRevB.110.104420,panigrahi2024spinchiralityfermionstirring}), and its implications remain to be seen. But there is some prior work suggesting that Eq.~(\ref{eq:Hspinchiral}) can lead to spin liquid physics when the Heisenberg term $J$ is antiferromagnetic \cite{gong_global_2017}.

\subsection{Some new things that I didn't talk about}

I will close these notes with a very lazy list of things that I didn't talk about, but which are interesting, relatively modern, and related to WCs.

\begin{enumerate}
	\item \emph{``Generalized'' WCs}: In moir\'e materials, the electrons experience a long(ish)-wavelength periodic potential in real space. At certain electron densities, the number of electrons is such that it can fill the potential in some regular, patterned way: one electron per potential well ($\nu = 1$), or one for every three wells $(\nu = 1/3)$, or one for every seven $(\nu = 1/7)$, etc. These states tend to get called ``generalized WCs'', since they require the long-ranged Coulomb interaction to be stable but are not the same as the jellium-model state with spontaneous symmetry breaking, as envisioned by Wigner. Here, unlike in the ``ungeneralized'' WC, the periodic potential does a lot of work in stabilizing the state. Experimentally, these special commensurate fillings correspond to an insulating (or more deeply insulating) state. See, e.g., Refs.~\cite{xu_correlated_2020, li_imaging_2021}.
	
	\item \emph{Bilayer WCs}: It is possible to arrange two 2DESs in close proximity to each other (separated by an insulating spacer layer), uncoupled by anything other than the Coulomb interaction. In this case, WCs can form in each layer when the densities are low, but they ``lock'' to each other in nontrivial ways to form different kinds of lattice patterns, such as a honeycomb lattice, square lattice, or rhombic lattice \cite{vilk_quasi-two-dimensional_1984, goldoni_stability_1996, esfarjani_bilayer_1995}. Layered materials provide a natural way to realize these kinds of situations (see, e.g., \cite{zhou_bilayer_2021} and \cite{zhou_electronic_2026}), and there are all sort of possibilities for new kinds of spin ordering \cite{esterlis2025magnetism}.
	
	\item \emph{Defect-induced magnetism}: Defects (interstitials and vacancies) in the WC can locally make a big change to the exchange pathways that produce magnetism by introducing tunneling pathways that have much lower tunneling action. Recent theory work has suggested that even a small concentration of such defects can dominate the magnetic ordering of a WC \cite{kim_interstitial-induced_2022}.
	
	\item \emph{Competition between WC and quantum Hall states}: What happens to the WC when one turns on a perpendicular magnetic field? Within the Einstein phonon picture, the magnetic field only deepens the WC state, since the Lorentz force acts to compress the electron wave packets. But experiments suggest that, if $r_s$ is not too large, the magnetic field can actually drive the WC state into an integer quantum Hall state when the Landau level filling factor $\nu = 2 \pi n \hbar /e B$ is very close to an integer (or to a special fraction like $1/3$, which produces a fractional quantum Hall state) \cite{falson_competing_2022}. This situation produces dramatic resistivity plots like this one, taken in ZnO \cite{falson_competing_2022}:
	
	\begin{figure}[H]
		\centering
		\includegraphics [width = 0.6\textwidth]{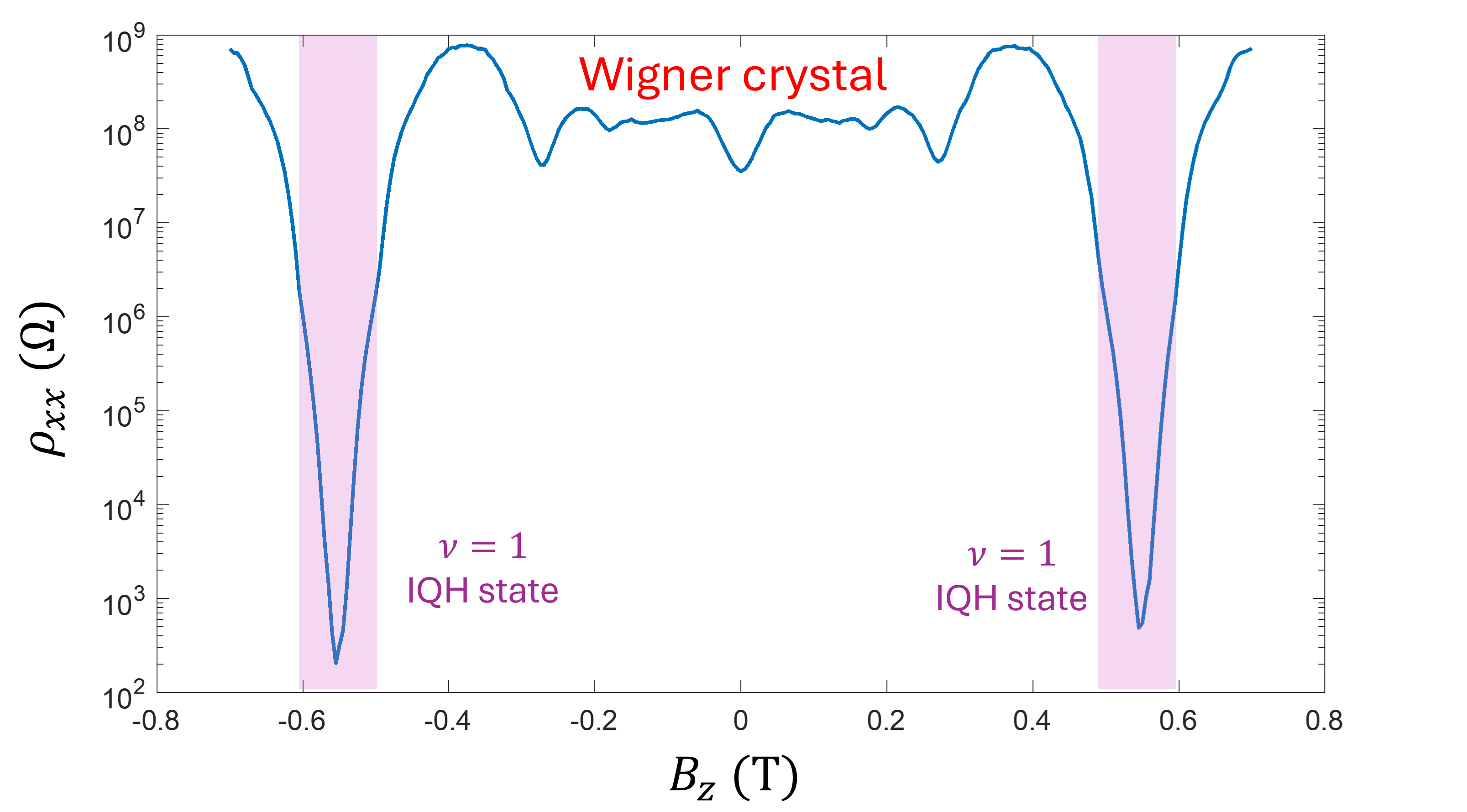}
		\caption{The longitudinal resistance $\rho_{xx}$ of a 2DES in ZnO \cite{falson_competing_2022} is plotted as a function of a perpendicular magnetic field $B_z$. The electron density for this curve corresponds to $r_s \approx 33$, and at $B=0$ the state resembles a WC, with large resistance and strong pinning behavior in the $I$-$V$ curve (see Fig.~\ref{fig:pinning}). However, when the magnetic field is such that $\nu \approx \pm 1$, the resistance abruptly dives down by over 6 orders of magnitude, indicating the formation of an integer quantum hall (IQH) state, before recovering back to the highly-insulating WC state. } 
		\label{fig:ZnO-WC}
	\end{figure}
	
	There is some very recent theory work offering a description of this phenomenon \cite{reddy_quantum_2026}. At asymptotically large $r_s$, the WC is stable at any value of magnetic field.
\end{enumerate}

\newpage

{\small
\bibliography{reference.bib}
\bibliographystyle{utphys}
}

\end{document}